\documentclass[12pt]{article}
\newcommand\bearst{\begin{eqnarray*}}
\newcommand\eearst{\end{eqnarray*}}
\def\half{\frac{1}{2}}

\def\half{\frac{1}{2}}
\newcommand{\slpe}{\raise.15ex\hbox{$/$}\kern-.57em\hbox{$p$}}
\newcommand{\slpartial}{\raise.15ex\hbox{$/$}\kern-.57em\hbox{$\partial$}}
\newcommand{\slp}{\raise.15ex\hbox{$/$}\kern-.57em\hbox{$p$}}
\newcommand{\slq}{\raise.15ex\hbox{$/$}\kern-.57em\hbox{$q$}}
\newcommand{\slk}{\raise.15ex\hbox{$/$}\kern-.57em\hbox{$k$}}
\newcommand{\sla}{\raise.15ex\hbox{$/$}\kern-.57em\hbox{$a$}}
\newcommand{\slA}{\raise.15ex\hbox{$/$}\kern-.57em\hbox{$A$}}
\newcommand{\slB}{\raise.15ex\hbox{$/$}\kern-.57em\hbox{$B$}}
\newcommand{\slD}{\raise.15ex\hbox{$/$}\kern-.57em\hbox{$D$}}
\newcommand{\slb}{\raise.15ex\hbox{$/$}\kern-.57em\hbox{$b$}}
\newcommand{\slc}{\raise.15ex\hbox{$/$}\kern-.57em\hbox{$c$}}
\newcommand{\sld}{\raise.15ex\hbox{$/$}\kern-.57em\hbox{$d$}}
\newcommand{\slW}{\raise.15ex\hbox{$/$}\kern-.57em\hbox{$W$}}
\newcommand{\slP}{\raise.15ex\hbox{$/$}\kern-.57em\hbox{$P$}}
\newcommand{\be}{\begin{equation}}
\newcommand{\ee}{\end{equation}}
\newcommand{\bear}{\begin{eqnarray}}
\newcommand{\ear}{\end{eqnarray}}
\newcommand{\ba}{\begin{eqnarray*}}
\newcommand{\ea}{\end{eqnarray*}}
\renewcommand{\theequation}{\arabic{section}.\arabic{equation}}

\newcommand{\no}{\noindent}

\begin{document}

\title{Two-Dimensional Thermofield Bosonization} 

\author{R. L. P. G. Amaral and L. V. Belvedere\\
\it{Instituto de F\'{\i}sica}\\
\it{Universidade Federal Fluminense}\\
\it{Av. Litor\^anea S/N, Boa Viagem, Niter\'oi, CEP. 24210-340}\\
\it{Rio de Janeiro - Brasil}\\
K. D. Rothe\\
\it{Institut f\"ur Theoretische Physik}\\
\it{Universit\"at Heidelberg}\\
\it{Philosophenweg 16, D-69120 Heidelberg}\\
\it{Germany}}

\date{\today}
\maketitle
\begin{abstract}
The main objective of this paper is to obtain an operator realization for the 
bosonization of fermions in $1 + 1$ dimensions, at finite, non-zero 
temperature $T$. This is achieved in the framework of the real time 
formalism of Thermofield Dynamics. Formally the results parallel those 
of the $T = 0$ case. The well known two-dimensional Fermion-Boson 
correspondences at zero temperature are shown to hold also at finite
temperature. In order to emphasize the 
usefulness of the operator realization for handling a large class of 
two-dimensional quantum field-theoretic
problems, we contrast this global approach with the cumbersome calculation
of the fermion-current two-point function in the imaginary-time formalism and
real time formalisms. The calculations also illustrate the very different ways
in which the transmutation from Fermi-Dirac to Bose-Einstein statistics is
realized.

\end{abstract}
\newpage
\section{Introduction}

The bosonization of fermions has proven in the past to be a very useful 
technique for solving quantum field theoretic models in 1+1 
dimensions \cite{AAR}. Recently this Fermion-Boson mapping has been
discussed for the case of a finite Temperature, using the imaginary
time formalism \cite{Delepine}, and the 
equivalence of the massive Thirring model and the sine-Gordon theory
was shown  also to hold at finite temperature. The main purpose of the 
present paper is to fill a gap in the literature, by considering  the 
operator formulation of two-dimensional
bosonization at finite temperature. To this end we shall use a real time
formalism  of Thermofield Dynamics \cite{U,Ojima,Mats,Das} . This will enable us to give 
a unified treatment of whole class of two-dimensional models at 
finite temperature.

The paper is organized as follows: We begin in section 2 by computing the 
two-point function of the current of free fermions within the imaginary time and real time formalisms. In this computation we show 
how the transmutation from Fermi-Dirac statistics to Bose-Einstein statistics is realized
in quite  different ways. These calculations  also serve to 
emphasize later the simplifying aspects of the operator thermofield 
bosonization.

In section 3 we consider the two-dimensional free massless scalar 
thermofield. The 
corresponding two-point function exhibits two infrared singularities, one 
similar to the zero temperature case
and a new temperature dependent one, implying an indefinite-metric Hilbert 
space on which the massless scalar thermofield acts. We then show that one
can nevertheless define positive definite Wick ordered exponentials of a
massless scalar thermofield, provided we associate with them a conserved 
charge (superselection rule). 

In section 4 we consider the two-dimensional Fermi thermofield and  
compute the corresponding two-point function. In the zero temperature limit 
we recover the known results.

In section 5 we then use the Wick-ordered exponentials of the 
free massless scalar thermofield as building blocks for the 
operator thermofield bosonization of the free massless Fermi field and
show that the Fermi thermofield satisfies the correct statistics. We also 
compute the fermion thermofield current
by a point-splitting limit and verify that the current satisfies the 
usual bosonization correspondence. As a matter of
fact, the well known Fermion-Boson correspondences at zero 
temperature 

\be
\bar \psi\,i\,\gamma^\mu\,\partial_\mu\,\psi\, \to\,
\frac{1}{2}\,:\partial_\mu \phi\,\partial^\mu \phi :\,,
\ee

\be\label{curbos}
\bar \psi\,\gamma^\mu\,\psi\, \to\,
-\,\frac{1}{\sqrt \pi}\,\epsilon^{\mu \nu}\,\partial_\nu \phi\,,
\ee

\be
M\,\bar \psi\,\psi\, \to\,
\,-\,\frac{\mu\,M}{\pi}:\cos 2 \sqrt \pi \phi :\,,
\ee

\no are shown to hold also at finite temperature.

In section 6 we illustrate the use of  
thermofield bosonization by solving the massless Thirring model at finite
temperature and conclude
with comments and outlook in section 7. We leave to the appendix the 
computation of the two-point function of the free massive scalar 
thermofield, and show that the massless scalar thermofield two-point function
is recovered in the zero mass limit.

\section{The Transmutation of  Fermi-Dirac to Bose-Einstein Statistics}

In this section we illustrate 
how the transmutation from Fermi-Dirac statistics to Bose-Einstein 
statistics is realized in the imaginary and real time formalisms in quite
different ways. These calculations  also serve to emphasize later 
the simplifying aspects of the operator thermofield bosonization.

\subsection{Current 2-point function in the imaginary time formalism}
In the following it is convenient to work in light-cone coordinates
$x_\pm = x_1 \pm ix_2$.  Defining correspondingly
$j_\pm = j_1 \pm ij_2$,
we have for the {\it euclidean} two-point current correlator at {zero} temperature 

\ba\label{current-correlator}
<j_+(x)j_+(y)> &=& -tr\left(\gamma_+iS_E(x-y)\gamma_+iS_E(y-x)\right)\nonumber
\\
&=&\int \frac{d^2k}{(2\pi)^2}\Pi^{(E)}_{++}(k)e^{ik\cdot(x-y)}
\ea

\no where the overall minus sign arises from the Fermi statistics, and
where, after performing the trace using  $\slq = \frac{1}{2}(\gamma_+q_- + \gamma_-q_+)$ and  $\gamma_+^2 = \gamma_-^2 = 0$,

\be\label{EPi++}
\Pi^{(E)}_{++}(k)= 4\int \frac{d^2p}{(2\pi)^2}
[(p-k)_1 + i(p-k)_2][ p_1 + ip_2] D_E(p-k)D_E(p)
\ee

\no with

\[
D_E(q) = \frac{1}{q^2} = \frac{1}{(q_1 + iq_2)(q_1 - iq_2)}\,.
\]

\no At finite temperature (\ref{current-correlator}) is replaced by

\[
<j_+(x)j_+(y)>_\beta = \sum_\ell\frac{1}{\beta}
\int \frac{dk_1}{2\pi}e^{ik_1(x_1-y_1)}
e^{i\omega_\ell(x_2 - y_2)}\tilde\Pi^{(E)}_{++}(k_1,\omega_\ell)
\]

\no where $\omega_\ell$ are the Matsubara frequencies 
$\omega_\ell = \frac{2\pi\ell}{\beta}$, with $\beta = 1/T$ the inverse of the
absolute temperature, and

\[
\tilde\Pi^{(E)}_{++}(k_1,\omega_\ell) = 4\sum_m \frac{1}{\beta}\int \frac{dp_1}{2\pi}
\frac{1}{[p_1 - i(m+\half)\frac{2\pi}{\beta}]
[p_1-k_1 - i(m-\ell+\half)\frac{2\pi}{\beta}]}\,.
\]

\no The evaluation of this 
expression requires regularization. Indeed, as is well known, formally 
the result depends on the order in which the $p_1$ integration and 
summation over $m$ is being done.

\bigskip\noindent
{\it Perform first the summation over $m$}

\bigskip
It is convenient to rewrite this expression as

\[
\tilde\Pi^{(E)}_{++}(k_1,\omega_\ell) = 4\left(\frac{i\beta}{2\pi }\right)^2
\sum_m \frac{1}{\beta}\int \frac{dp_1}{2\pi}\frac{1}{[m+ \half + i\frac{\beta}{2\pi}p_1]
[m +\half + i\frac{\beta}{2\pi}(p_1 - k_1+i\frac{2\pi}{\beta}\ell)]}
\]

\no Making use of the familiar formula \cite{Das}

\[
\sum_m \frac{1}{[(m+\half) + ix]}\frac{1}{[(m+\half)+iy]}=
\frac{\pi}{x-y}(\tanh\pi x - \tanh\pi y)
\]

\no we obtain

\bearst
\tilde\Pi^{(E)}_{++}(k_1,\omega_\ell) = -\frac{2}{k_1-i\frac{2\pi\ell}{\beta}}
\int \frac{dp_1}{2\pi} \left(\tanh\frac{\beta}{2}p_1 - \tanh\frac{\beta}{2}(p_1 - k_1)\right)\,.
\eearst

\no Now

\[
\tanh\frac{\beta}{2}q_1 = \epsilon(q_1)(1 - 2N_F(|q_1|))
\]

\no where $N_F(|q_1|)$ is the Fermi-Dirac distribution

\be\label{Fermi-Dirac}
N_F(|q_1|) = \frac{1}{e^{\beta|q_1|}+1}\,.
\ee 

\no Proceeding with the $p_1$ integration, one finds

\[
\tilde\Pi^{(E)}_{++}(k_1,\omega_\ell)= -\frac{2}{\pi}\frac{k_1}{k_1 -i\frac{2\pi\ell}{\beta}}
\]

\no A similar result is obtained in this way in the $T=0$ case \cite{Nielsen}.

\bigskip\noindent
{\it Perform first the integral over $p_1$}

\bigskip\noindent
Performing first the integral in $p_1$, one obtains 

\bearst
&&\int \frac{dp_1}{2\pi}
\frac{1}{[p_1 - i(m+\half)\frac{2\pi}{\beta}]
[p_1-k_1 - i(m-\ell+\half)\frac{2\pi}{\beta}]}=\\
&&=-i\left[\theta(2m+1)\theta(-2m+2\ell-1)
- \theta(-2m-1)\theta(2m-2\ell+1)\right]\frac{1}{k_1 - i\frac{2\pi\ell}{\beta}}
\eearst

\no Summing now over $m$ one finds,

\bearst
\tilde\Pi^{(E)}_{++}(k_1,\omega_\ell) &=& 
\frac{-4i}{\beta}(\theta(\ell) + \theta(-\ell - 1))
\frac{\ell}{k_1 - i\frac{2\pi\ell}{\beta}}\\
&=&-\frac{2}{\pi}\frac{i\frac{2\pi\ell}{\beta}}{k_1 - i\frac{2\pi\ell}{\beta}}
\eearst

\no This is again the analogue of the corresponding 
$T=0$ result \cite{Nielsen}.

Averaging now the results we obtain

\be\label{Pitilde++}
\tilde\Pi_{++}(k_1,\omega_\ell)= -\frac{1}{\pi}\frac{k_1 + i\frac{2\pi\ell}{\beta}}
{k_1 -i\frac{2\pi\ell}{\beta}}
\ee

\no As the final step it remains to compute

\[
<j_+(x)j_+(y)>_\beta = -\frac{1}{\pi}\sum_\ell\frac{1}{\beta}\int\frac{dk_1}{2\pi}e^{ik_1(x_1-y_1)}
e^{i\frac{2\pi\ell}{\beta}(x_2 - y_2)}\frac{k_1 + i\frac{2\pi\ell}{\beta}}
{k_1 - i\frac{2\pi\ell}{\beta}}\,.
\]

\no We may write this as

\[
<j_+(x)j_+(y)>_\beta = -\frac{1}{\pi}\partial^2_+ D_E^{(\beta)}(x-y)
\]

\no where 

\[ 
D_E^{(\beta)}(z)= \sum_\ell\frac{1}{\beta}\int \frac{dk_1}{2\pi}\frac{e^{ik_1z_1}
e^{i \frac{2\pi\ell}{\beta}z_2}}{(k_1^2 + \frac{2\pi\ell}{\beta})^2}
\]

\no The result, valid for $|z_2|< \beta$ is well known. Replacing the sum by an integral by using a type of Sommerfeld-Watson transform, one finds,  

\[
D_E^{(\beta)}(z)=\int\frac{d^2k}{(2\pi)^2}e^{ik\cdot z}
\left[\frac{1}{k^2} -i 2\pi\delta(k^2)N_B(|k_1|)\right] 
\]

\no where $N_B(|k_1|)$ is now the {\it Bose-Einstein} distribution function

\be\label{Bose-Einstein}
N_B(|k_1|) = \frac{1}{e^{\beta{|k_1|}}-1}\,.
\ee

\no Continuing to Minkowski space this is just the finite-temperature
scalar two point function of the real-time formalism. 
\be\label{DB}
D^{(\beta)}_E(z) \to iD^{(\beta)}_B(z) = 
i\int\frac{d^2k}{(2\pi)^2}e^{-ik\cdot z}
\left[\frac{1}{k^2+i\epsilon} + 2\pi i\delta(k^2)N_B(|k_1|)\right]
\ee
so that
\be\label{jj}
<Tj_+(x)j_+(y)>_\beta = \frac{1}{\pi}\partial^2_+ iD_B^{(\beta)}(x-y)
\ee

 We have thus
verified the correspondence (\ref{curbos}) in the imaginary-time formalism,
witnessing the metamorphosis from the Fermi-Dirac distribution describing
the intermediate fermionic states to be Bose-Einstein distribution
in the equivalent bosonic description!
The demonstration for the $--$ component of the current correlator proceeds in 
the same way. Since the $+-$ components vanish because $\gamma_{\pm}^2 = 0$,
this proves the full equivalence.

We now show that in a quite different, and surprising way, this 
metamorphosis is realized in the real-time formalism.

\subsection{Current 2-point function in the real time formalism}

Using the standard rules of the real-time formalism, we have for the
++ component of the current two-point function, 

\[
<Tj_{++}(x)j_{++}(y)>_\beta = -tr\left(\gamma_+iS_F^{(\beta)}(x-y)\gamma_+iS_F^{(\beta)}(y-x)\right)\,,
\]

\no where

\[
S_F^{(\beta)}(z) = \int d^2q \,{\slq} D_F^{(\beta)}(q)e^{-iq\cdot z}
\]

\no with

\be\label{scalar-2pt-function}
 D_F^{(\beta)}(q)= \frac{1}{q^2 + i\epsilon}
+ 2\pi i \delta(q^2) N_F(|\vec q|)\,,
\ee

\no where $ N_F(|\vec q|)$ is the Fermi-Dirac distribution (\ref{Fermi-Dirac}). Here we use
$x^{\pm}=x^0\pm x^1$ and correspondingly $\gamma^\pm=\gamma^0\pm\gamma^1$.
Noting again that $\slq = \frac{1}{2}(\gamma_+q_- + \gamma_-q_+)$ and 
$\gamma_+^2 = \gamma_-^2 = 0$, we find

\be\label{j+j+}
<Tj_+(x)j_+(y)>_\beta = 
\int \frac{d^2k}{(2\pi)^2}\Pi_{++}(k)e^{-ik\cdot(x-y)}
\ee

\no with

\be\label{Pi++}
\Pi_{++}(k)= 4\int \frac{d^2p}{(2\pi)^2}
(p-k)_+ p_+ D_F^{(\beta)}(p-k)D_F^{(\beta)}(p)\,.
\ee

\no We next compute explicitly the integral in (\ref{Pi++}). Separating 
into terms independent of the temperature ($\Pi^{(0)}_{++}$), linear in the 
Bose-Einstein distribution ($\Pi^{(1)}_{++}$), and 
quadratic in $N_F(|\vec q|)$ ($\Pi^{(2)}_{++}$), we have 

\[
\Pi_{++}(k) = \Pi^{(0)}_{++}(k) + \Pi^{(1)}_{++}(k) + \Pi^{(2)}_{++}(k)
\]

\no we have

\bearst
 \Pi^{(0)}_{++}(k) &=&  4\int \frac{d^2p}{(2\pi)^2}
\frac{(p-k)_+ p_+ }{[(p-k)^2 + i\epsilon](p^2 + i\epsilon)}\\
 \Pi^{(1)}_{++}(k) &=&  4\int \frac{d^2p}{(2\pi)^2}
[\frac{(p-k)_+ p_+ }{(p-k)^2 + i\epsilon}(2\pi i) N_F(|p_1|)\delta(p^2)\\
&+& \frac{(p-k)_+ p_+ }{p^2 + i\epsilon}(2\pi i) N_F(|p_1-k_1|)\delta((p-k)^2)]\\
\Pi^{(2)}_{++}(k) &=&  4\int \frac{d^2p}{(2\pi)^2}(p-k)_+ q_+ 
(2\pi i) N_F(|p_1|)\delta(p^2)(2\pi i) N_F(|p_1-k_1|)\delta((p-k)^2)\,.
\eearst

\no $\Pi^{(0)}_{++}(k)$ can be calculated in a variety of ways. A regularization consistent with the conservation law $\partial_\mu j^\mu = 0$ yields

\be\label{Pi(0)++}
\Pi^{(0)}_{++}(k) = -\frac{i}{\pi}\frac{k_+k_+}{k^2}\,.
\ee

\no We now make use of the Plemelj's decomposition

\[
\frac{\alpha}{\alpha(x-x_0) + i\epsilon} =
{\cal P}\left(\frac{1}{x-x_0}\right) -i\pi\epsilon(\alpha)\delta(x-x_0)
\]

\no in order to rewrite $\Pi^{(1)}_{++}(k)$ and  $\Pi^{(2)}_{++}(k)$ in the form

\ba\label{Pi123}
&&\Pi^{(1)}_{++}(k) =  4\int \frac{d^2p}{(2\pi)^2}
[{\cal P}\left(\frac{1}{(p-k)_-}\right)
(2\pi i) N_F(|p_1|)p_+\delta(p^2)\\
&&+{\cal P}\left(\frac{1}{p_-}\right) 
(2\pi i) N_F(|p_1-k_1|)(p-k)_+\delta((p-k)^2)]\nonumber\\
&&- 4\int \frac{d^2p}{(2\pi)^2}(2\pi i)\left[\epsilon(p_+)\delta(p_-)
\epsilon(p_+ -k_+)\delta((p - k)_-)[N_F(|p_1|)+N_F(|p_1-k_1|)\right]\nonumber
\ea

\no and

\[
\Pi^{(2)}_{++}(k) = 4\int \frac{d^2p}{(2\pi)^2}
(2\pi i) N_F(|p_1|)\epsilon(p_+)\delta(p_-)
(2\pi i) N_F(|p_1-k_1|)\epsilon(p_+ -k_+)\delta((p - k)_-)\,.
\]

\no Now,

\bearst
{\cal P}\left(\frac{1}{(p-k)_-}\right)(2\pi i) 
N_F(|p_1|)p_+\delta(p^2)&=&
-\frac{1}{k_-}N_F\left(\frac{|p_+|}{2}\right)\epsilon(p_+)\delta(p_-)\\
{\cal P}\left(\frac{1}{p_-}\right)(2\pi i) N_F(|p_1-k_1|)(p-k)_+\delta((p-k)^2)&=&
\frac{1}{k_-}N_F\left(\frac{|(p-k)_+|}{2}\right)
\epsilon(p_+ - k_+)\delta((p-k)_-)\,.
\nonumber
\eearst

\no  Since the left hand side of the above expressions are odd functions of $p_+$ and $p_+ - k_+$, respectively, they do not contribute to the integral above. In terms of light-cone
coordinates we have $d^2p = \frac{1}{2}dp_+dp_-$. Performing the
$p_-$ integration we are left with 

\ba\label{Pi1++}
\Pi_{++}(k)=&& -\frac{i}{\pi}\frac{k_+k_+}{k^2}
- 4(\pi i)^2\delta(k_-)\int \frac{dp_+}{(2\pi)^2}
\epsilon(p_+-k_+)\epsilon(p_+)\\
&&\left[N_F\left(\frac{|p_+|}{2}\right) + N_F\left(\frac{|(p_+ -k_+)|}{2}\right) - 2N_F\left(\frac{|p_+|}{2}\right)N_F\left(\frac{|p_+-k_+|}{2}\right)\right]
\nonumber
\ea

\no We now make the change of variable
$\frac{p_+}{2} = q+\frac{k}{2}$ with $k=k_+/2$. Let us suppose for the moment that
$k_+$ is positive. We may then split the integral in the following way:

\ba\label{symPi++}
\Pi_{++}(k) &=& -\frac{i}{\pi}\frac{k_+k_+}{k^2}\\
&+&2\delta(k_-)\int_{-\infty}^{-\frac{k}{2}} dq 
\left[N_F(-q -\frac{k}{2}) + N_F(-q +\frac{k}{2})
-2N_F(-q -\frac{k}{2})N_F(-q +\frac{k}{2})\right]\nonumber\\
&+&2\delta(k_-)\int^{\infty}_{\frac{k}{2}} dq 
\left[N_F(q +\frac{k}{2}) + N_F(q -\frac{k}{2})
-2N_F(q +\frac{k}{2})N_F(q -\frac{k}{2})\right]\nonumber\\
&+&2\delta(k_-)\int_{-\frac{k}{2}}^\frac{k}{2} dq
\left[ N_F(q +\frac{k}{2}) + N_F(\frac{k}{2}-q)
- 2N_F(q +\frac{k}{2})N_F(\frac{k}{2}-q)\right]\nonumber
\ea

\no Making now explicit use of the Fermi-Dirac distribution (\ref{Fermi-Dirac}),
 we obtain from here

\ba
\Pi_{++}(k) &=& -\frac{i}{\pi}\frac{k_+k_+}{k^2}\\
&+&4\delta(k_-)\left(\int^{\infty}_{\frac{k}{2}}dq 
\frac{\cosh\beta\frac{k}{2}}{\cosh\beta\frac{k}{2} + \cosh\beta q}
- \int_0^{\frac{k}{2}}dq 
\frac{\cosh\beta q}{\cosh\beta\frac{k}{2} + \cosh\beta q}\right)\nonumber 
\ea

\no which can be rewritten in the form

\be
\Pi_{++}(k) = -\frac{i}{\pi}\frac{k_+k_+}{k^2}+
2\delta(k_-)\left[I(k)\cosh\beta\frac{k}{2}-k\right]
\ee

\no where

\bearst
I(k) &=& \int_{-\infty}^\infty dq 
\frac{1}{\cosh\frac{\beta k}{2} + \cosh\beta q}\\
&=& \frac{k}{\sinh\beta\frac{k}{2}}
\eearst

Putting things together we thus have

\[
\Pi_{++}(k) = -\frac{i}{\pi}\frac{k_+k_+}{k^2}+
2k_+\delta(k_-)\frac{1}{e^{\beta\frac{k_+}{2}}-1}
\]

\no Our analysis was done for $k_+ \ge 0$.
Since our original expression (\ref{symPi++}) is symmetric under $k\to -k$, we find that in general

\[
\Pi_{++}(k) = -\frac{i}{\pi}\frac{k_+k_+}{k^2}+
2|k_+|\delta(k_-)\frac{1}{e^{\beta\frac{|k_+|}{2}}-1}
\]

\no Noting that

\[
|k_+|\delta(k_-) = k_+k_+\delta(k^2)
\]

\no we finally conclude that

\[
\Pi_{++}(k) = -\frac{i}{\pi}k_+k_+\left(\frac{1}{k^2} 
+ 2\pi i\delta(k^2)N_B(|k_1|)\right)
\]

\no where $N_B(|k_1|)$ is now the {\it Bose-Einstein} 
distribution (\ref{Bose-Einstein}). A corresponding result is found for $\Pi_{--}(k)$,
while again $\Pi_{+-} = \Pi_{-+}=0$. We thus conclude that

\[
<Tj_\mu(x)j_\nu(y)>_\beta = \frac{1}{\pi}\tilde\partial_\mu\tilde\partial_\nu iD_B^{(\beta)}(x-y)
\]

\no in accordance with the correspondence (\ref{curbos}) for finite 
temperature. Notice that unlike in the case of the imaginary-time formalism, we have now witnessed the actual conversion of the Fermi-Dirac to the
Bose-Einstein distribution through an integration process.

\section{Free Massless Scalar Thermofield}
\setcounter{equation}{0}

As the first step we shall consider the 
massless scalar field since it is the building block
in the bosonization of two-dimensional quantum field theory models. The 
construction of a quantum field theory at finite temperature requires
doubling the number of fields degrees of freedom \cite{U,Ojima,Mats,Das}. This is achieved by
introducing fictitious ``tilde'' operators corresponding to each of the 
operators describing the system considered. This fictitious system is
an identical copy of the original system under consideration, except for the opposite norm of the corresponding field. This entails a
doubling of the Hilbert space.

To begin with, let us introduce the free scalar thermal doublet\footnote{The conventions used 
are: 

$$\gamma^0 = \pmatrix{0 & 1 \cr 1 & 0}\,,\gamma^1 
= \pmatrix{0 & 1 \cr - 1 & 0}\,,\gamma^5 = \gamma^0 \gamma^1\,\,,\,\,
\epsilon^{0 1} = 1\,,\,g^{00} = 1\,,\,x^\pm = x^0 \pm x^1\,,\,\partial_\pm 
= \partial_0 \pm \partial_1\,.
$$ 

\no For the free
massless scalar field $\phi (x) = \phi (x^-) + \phi (x^+)$, and for the pseudo-scalar
field $\varphi (x) = \varphi (x^-) - \varphi (x^+)$.},

\be
\Phi = \pmatrix{\phi \cr \widetilde \phi}\,.
\ee

\no corresponding to the total Lagrangian density,

\be
{\cal L}_T = {\cal L} - \widetilde{\cal L} = \frac{1}{2}\,\partial_\mu \phi \partial^\mu \phi\,-\,\frac{1}{2}\,\partial_\mu \widetilde \phi \partial^\mu \widetilde \phi \,,
\ee

\no with the corresponding equations of motion ($\partial_\pm = \partial_0 \pm \partial_1$),

\be\label{em}
\partial_+\,\partial_-\,\phi (x) = 0\,\,\,,\,\,\,\partial_+ \partial_- \widetilde \phi (x) = 0\,.
\ee

\no In view of the equations of motion (\ref{em}), the two-dimensional free massless scalar field can
be decomposed in terms of left- and right-movers  ($x^\pm = x^0 \pm x^1$),

\be
\phi (x) = \phi_{_{L}} (x^+) + \phi_{_{R}} (x^-)\,,
\ee

\no and similarly for $\widetilde\phi (x)$. In order to simplify the notation, we shall omit the subscripts, 
which are taken to be implied by the arguments $x^\pm$.

In the quantized theory, the two-dimensional 
Boson field at zero temperature is described by the field 
operator ,

\be
\phi (x) = \int_{- \infty}^{+ \infty}\,(dp)\,\Big (\,f_p (x)\,a (p^1)\,
+\,f^\ast_p (x)\,a^\dagger (p^1)\,
\Big )\,.
\ee

\be
f_p (x) = e^{- i p^\mu x_\mu}\,,
\ee

\be
(dp) = \frac{dp^1}{\sqrt{(2 \pi)(2 \vert p^1 \vert)}}\,.
\ee

\no The tilde conjugation is defined by the property

\be
\widetilde{(ca)} = c^\ast \widetilde a\,,
\ee

\no such that the tilde conjugated field is given by,

\be
\widetilde \phi (x) = \int_{- \infty}^{+ \infty}\,(dp)\,
\Big (\,f^\ast_p (x)\,\widetilde a (p^1)\,+\,f_p (x)\,
\widetilde a^\dagger (p^1)\,
\Big )\,.
\ee

\no At $T = 0$ these fields are independent,

\be
[ \phi (x)\,,\,\widetilde \phi (y) ]\,=\,0\,.
\ee

The Fock vacuum state is,

\be
\vert  \widetilde 0 , 0 \rangle = \vert \widetilde 0 \rangle \otimes \vert 0 \rangle\,,
\ee

\no with the property,

\be
a (p^1)\,\vert 0 \rangle = 0\,\,\,,\,\,\,\widetilde a (p^1)\,\vert \widetilde 0 \rangle = 0\,.
\ee

\no Decomposing the free massless fields 
into left- and right-components, we obtain 

\be\label{phi}
\phi (x^\pm) = \int_0^\infty\,\frac{dp^1}{ \sqrt{4 \pi \vert p^1 \vert}}\,
\Big [\,f_p (x^\pm)\,\pmatrix{a (-p^1) \cr a (p^1)}\,
+\,f^\ast_p (x^\pm)\,\pmatrix{a^\dagger (-p^1) \cr a^\dagger (p^1)}\,\Big ]\,,
\ee

\be\label{phit}
\widetilde \phi (x^\pm) = \int_0^\infty\,\frac{dp^1}{\sqrt{4\pi \vert p^1\vert}}\,
\Big [\,f^\ast_p (x^\pm)\,\pmatrix{\widetilde a (-p^1)\cr \widetilde a (p^1)}\,
+\,f_p (x^\pm)\,\pmatrix{\widetilde a^\dagger (-p^1)\cr \widetilde a^\dagger (p^1)}\,\Big ]\,,
\ee

\no where,

\be
f_p (x^\pm) = e^{\,-i\,p\, x^\pm }\,,
\ee

\no and

\be
[ \phi (x^\pm)\,,\,\phi (y^\mp) ] 
= [ \widetilde\phi (x^\pm)\,,\,\widetilde\phi (y^\mp) ] = 0 \,.
\ee

\no Note that in Eqs. (\ref{phi}) and (\ref{phit}) the limits 
of integrations are $[ 0 , \infty)$.

In Thermofield Dynamics
the temperature-dependent vacuum is defined by,

\be
\vert 0(\beta)   \rangle = U_B (\theta_B ) \vert 0, \widetilde 0 \rangle\,,
\ee

\no where the unitary operator $U_B (\theta_B )$ is given by,

\be
U_B (\theta_B ) = e^{ - i {\cal Q} (\theta_B)} = 
e^{\,- \int_{- \infty}^{+ \infty} dp^1 \Big (\widetilde a (p^1)  a (p^1) - 
a^\dagger (p^1) \widetilde a^\dagger (p^1) \Big )
\theta_B (\vert p^1\vert ,\beta )}\,,
\ee

\no and the Bogoliubov parameter $\theta_B(\vert p^1 \vert, \beta )$ is
implicitly defined by

\be
\sinh \theta_B (\vert p^1 \vert ; \beta )\,=\,
\frac{e^{\,-\,\beta \vert p^1 \vert /2}}
{\sqrt{1 - e^{\,-\beta \vert p^1 \vert }}}\,,
\ee

\be
\cosh \theta_B (\vert p^1 \vert ; \beta )\,=\,\frac{1}{\sqrt{1 - e^{\,-\beta 
\vert p^1 \vert}}}\,,
\ee

\no with the Bose-Einstein statistical weight is given by,

\be
N_B (\vert p^1 \vert ; \beta ) = \sinh^2 \theta_B (\vert p^1 \vert ; \beta )\,=\,
\frac{1}{e^{\beta \vert p^1 \vert} - 1}\,.
\ee

\no The correspondingly transformed annihilation operators are given by,

\be
a (p^1 ; \beta ) = U_B ( - \theta_B  )\, a (p^1) \,U_B (\theta_B ) = a (p^1) 
\cosh \theta_B (\vert p^1 \vert ; \beta ) -
\widetilde a^\dagger (p^1) \sinh \theta_B (\vert p^1\vert ; \beta )\,,
\ee

\be
\widetilde a (p^1 ; \beta ) = U_B ( - \theta_B ) \widetilde a (p^1) U_B (\theta_B ) = 
\widetilde a (p^1) \cosh \theta_B (\vert p^1\vert ; \beta ) -
a^\dagger (p^1) \sinh_B \theta (\vert p^1\vert ; \beta )\,.
\ee

The vacuum state at finite temperature satisfies

\be
a (p^1 ; \beta )\,\vert 0 (\beta ) \rangle\,=\,0\,,\;\;\;\;\;
\widetilde a (p^1 ; \beta )\,\vert 0 (\beta ) \rangle\,=\,0\,.
\ee

\no {\it Here, and in what follows}, $p$ {\it is defined to be} $p = \vert p^1 \vert$.

The thermofield operators, that act on the 
Fock vacuum $ \vert 0, \widetilde 0 \rangle$ are given by,

\be
\phi (x^\pm ; \beta ) = \phi^{(+)} ( x^\pm; \beta ) + \phi^{(-)} ( x^\pm; \beta )\,,
\ee

\be
\widetilde\phi (x^\pm ; \beta ) = \widetilde{\phi}^{(+)} ( x^\pm; \beta ) + 
\widetilde{\phi}^{(-)} ( x^\pm; \beta )\,,
\ee

\no with

\be
\phi^{(\pm)} ( x^\pm; \beta ) = \phi_c ^{(\pm)}( x^\pm; \beta ) - 
\widetilde \phi_s^{(\pm)} ( x^\pm; \beta )\,,
\ee

\be
\widetilde\phi^{(\pm)} ( x^\pm; \beta ) = \widetilde{\phi}_c ^{(\pm)}( x^\pm; \beta ) - 
\phi_s^{(\pm)} ( x^\pm; \beta )\,,
\ee

\be
\phi_c^{(\pm)} ( x^\pm; \beta ) = \int_0^\infty\,(d p)\,
\pmatrix{f_p (x^\pm)\, a(\mp p) \cr f^\ast_p (x^\pm)\, 
a^\dagger(\mp p)}\,\, \cosh \theta_B (p ; \beta )\,,
\ee

\be
\phi_s ^{(\pm)}( x^\pm; \beta ) = \int_0^\infty\,(d p)\,
\pmatrix{f_p (x^\pm)\,a(\mp p) \cr f^\ast_p (x^\pm)\, 
a^\dagger(\mp p)}\, \sinh \theta_B (p ; \beta )\,,
\ee

\be
\widetilde \phi_s ^{(\pm)}( x^\pm; \beta ) = \int_0^\infty\,(d p)\,
\pmatrix{f^\ast_p (x^\pm)\,\widetilde a(\mp p)\cr f_p (x^\pm)\, 
\widetilde a^\dagger(\mp p)}\, \sinh \theta_B (p ; \beta )\,,
\ee

\be
\widetilde\phi_c^{(+)} ( x^\pm; \beta ) = \int_0^\infty\,(d p)\,
\pmatrix{f^\ast_p (x^\pm)\,\widetilde a(\mp p) \cr f_p (x^\pm)\, 
\widetilde a^\dagger(\mp p)}\, \cosh \theta_B (p ; \beta )\,.
\ee

\no The commutators of the thermofield are the same as that at zero temperature,

\be\label{ccr}
[ \phi (x^\pm ; \beta ) , \phi (y^\pm ; \beta ) ] = [ \phi_c(x^\pm; \beta ) , \phi_c ( y^\pm; \beta ) ] 
+ 
[ \widetilde \phi_s(x^\pm; \beta ) , \widetilde \phi_s ( y^\pm; \beta ) ]  =
 \,[ \phi (x^\pm) , \phi (y^\pm) ]\,,
\ee

\noindent so that
\be
[ \phi (x^\pm ; \beta ) , \dot\phi (y^\pm ; \beta ) ] =
 \,[ \phi (x^\pm) , \dot\phi (y^\pm) ]\,,
\ee

\no and

\be
[ \phi (x^\pm ; \beta ) , \widetilde \phi (y^\pm ; \beta ) ] = 0\,,
\ee

As in the zero temperature case, the massless scalar themofield carries an 
identically conserved topological current associated with the Gauss' law

\be
J^\mu (x ;\beta ) = \partial_\nu  \epsilon ^{\mu \nu} \varphi (x ; \beta )\,,
\ee

\no and the field $\varphi (x ; \beta )$ is the dual 
of $\phi ( x ; \beta )$,

\be
\epsilon^{\mu \nu}\,\partial_\nu\,\varphi (x ; \beta ) = 
\partial^\mu\,\phi (x ; \beta )\,.
\ee

\no The corresponding charge is formally given by

\be
{\cal Q}_\beta = \int J^0 (x ; \beta)\,d x^1\,.
\ee

\subsection{Scalar Thermofield Two-point Function}

For free fields the $n$-point functions are determined by the two-point
function. The diagonal contribution to the Schwinger two-point function 
is given by,

\begin{eqnarray}
\langle 0(\beta ) \, \vert \phi (x^\pm) \,
\phi (y^\pm)\, \vert 0(\beta ) \rangle\,&=&\,
\langle 0, \widetilde 0 \vert \phi (x^\pm ; \beta ) 
\phi (y^\pm ; \beta ) \vert \widetilde 0, 0 \rangle \\
&=&[ \phi_c^{(+)} (x^\pm; \beta )\,,\,\phi_c^{(-)} (y^\pm; \beta) ]\,+\,
[ \widetilde \phi_s^{(+)} (x^\pm; \beta )\,,\,\widetilde \phi_s^{(-)} (y^\pm; \beta) ]\,
\\
&=&
D^{(+)}_c (x^\pm - y^\pm; \beta ) + \widetilde D^{(+)}_s (x^\pm - y^\pm; \beta )\,,
\end{eqnarray}

\no where in our simplified notation,

\be
D^{(+)}_c (x^\pm - y^\pm; \beta )\,=\,\int_0^\infty\,\frac{dp}{(2 \pi )(2 p)}\,f_p (x^\pm)\,
f^\ast_p (y^\pm)\,\cosh^2 \theta_B (p; \beta )\,,
\ee

\be
\widetilde D^{(+)}_s (x^\pm - y^\pm; \beta )\,=\,\int_0^\infty\,\frac{dp}{(2 \pi )
(2 p)}\,f^\ast_p (x^\pm)\,
f_p (y^\pm)\,\sinh^2 \theta_B (p; \beta )\,.
\ee

\no We get,

\be\label{2pf}
\langle 0, \widetilde 0 \vert \phi (x^\pm ; \beta ) 
\phi (y^\pm ; \beta ) \vert \widetilde 0, 0 \rangle =
D^{(+)}_o (x^\pm - y^\pm)+\frac{1}{2\pi}\,\int_0^\infty\,\frac{dp}{p}\,[\cos p (x^\pm - y^\pm) ]\,N_B (p; \beta ),
\ee

\no where the zero temperature two-point function is given by,

\be
D^{(+)}_o (x^\pm - y^\pm) = -\,\frac{1}{4 \pi}\,\ell\mathit{n} [i \mu ( x^\pm - y^\pm -i \epsilon)]\,.
\ee

\no where $\mu$ is an arbitrary infrared (IR) regulator. Besides the 
usual infrared 
divergence, the integral
in (\ref{2pf}) introduces an additional divergence. The integral 
appearing in (\ref{2pf}) is given by, \footnote{Gradshteyn - 4 Edition, pg. 495, 19 $m = 1$;

$$
\int_0^\infty\,\frac{\cos (a x) - \cos (b x)}{e^{\beta x} - 1}\,\frac{dx}{x}\,=\,
\frac{1}{2}\,\ln\,\frac{a \sinh \frac{b \pi}{\beta}}{b \sinh \frac{a \pi}{\beta}}\,.
$$
},

$$
\frac{1}{2\pi}\,\int_0^\infty\,\frac{dp}{p}\,[\cos p (x^\pm - y^\pm) ]\,N_B (p; \beta )
$$

\be\label{int}
=\,\frac{1}{4 \pi} \ln (x^\pm - y^\pm)\,-\,\frac{1}{4 \pi} \ln \Bigg \{ \sinh \frac{\pi (x^\pm - y^\pm)}{\beta } \Bigg \}\,
-\,\frac{1}{4 \pi} \ln \frac{\beta}{\pi}\,+\,
\frac{1}{2 \pi}\,\int_0^\infty\,\frac{dp}{p\,(e^{\,\beta p} - 1)}\,.
\ee

\no The integral appearing in (\ref{int}) is infrared divergent. We shall 
introduce an infrared cut-off $\mu^\prime$ and define

\be\label{z}
\mathit{z} (\beta, \mu^\prime )\,=\,\int_{\mu^\prime}^\infty\,\frac{dp}{p\,(e^{\,\beta p} - 1)}
\,=\,\int_{\mu^\prime}^\infty\,\frac{dp}{p}\,N_B (p; \beta )\,,
\ee

\no which corresponds to the mean number of particles having 
momenta in the range $[{\mu^\prime}, \infty ]$,

\be
\mathit{z} (\beta, {\mu^\prime} )\,=\,\langle 0 ; \beta \, \vert \,{\cal N}_B \, 
\vert\,0(\beta ) \rangle\,,
\ee

\no where

\be
{\cal N}_B \,=\,\int_{\mu^\prime}^\infty\,\frac{d p}{p}\,a^\dagger (p)\,a (p)\,,
\ee

\no is the operator for the number of particles. For a fixed temperature 
the asymptotic behavior of the integral (\ref{z}), besides the 
logarithmic singularity, exhibits also an algebraic singularity, 

\be
\mathit{z} ({\mu^\prime} \approx 0 ; \beta ) \rightarrow  \frac{1}{\beta\,{\mu^\prime}}
 + \frac{1}{2}\,\ln (\beta {\mu^\prime} )\,.
\ee

\no As we shall see, the 
integral (\ref{z}) plays a crucial role in determining the selection rule for the 
correlation functions of Wick exponentials of
the free massless scalar thermofield.

From (\ref{2pf}), (\ref{int}) and (\ref{z}) the 
two-point function for the left- and right-movers seen to be given by ,

\be\label{btpf}
\langle 0, \widetilde 0 \vert \phi (x^\pm ; \beta ) \phi (y^\pm ; \beta ) \vert \widetilde 0, 0 \rangle 
\,=\,
-\,\frac{1}{4 \pi} \ln \Bigg \{\,i\,\mu\,\frac{\beta}{\pi}\, 
\sinh \frac{\pi (x^\pm - y^\pm-i\,\epsilon )}{\beta } \Bigg \}\,+\,
\frac{1}{2 \pi}\,\mathit{z} (\beta , {\mu^\prime} )\,.
\ee

For the field $\widetilde \phi (x^\pm)$, we obtain

\be\label{tbtpf}
\langle 0, \widetilde 0 \vert \widetilde \phi (x^\pm ; \beta ) \widetilde \phi (y^\pm ; \beta ) \vert \widetilde 0, 0 \rangle 
\,=\,
-\,\frac{1}{4 \pi} \ln \Bigg \{\,i\,\mu\,\frac{\beta}{\pi}\, 
\sinh \frac{\pi (x^\pm - y^\pm+i\,\epsilon )}{\beta } \Bigg \}\,-\,\frac{i}{4}\,+\,
\frac{1}{2 \pi}\,\mathit{z} (\beta , {\mu^\prime} )\,.
\ee

\no The limit $\epsilon \rightarrow 0^+$ was introduced as a 
prescription to properly define the zero-temperature
limit of the above expression as a tempered distribution. For the space-time contribution, the short-distance 
limit $x \approx y$ coincides with the $T \rightarrow 0$ limit,

\be
\langle 0, \widetilde 0 \vert \phi (x^\pm ; \beta ) 
\phi (y^\pm ; \beta ) \vert \widetilde 0, 0 \rangle\,
\rightarrow\,-\,\frac{1}{4 \pi} \ln \Big \{\,i\,\mu\,(x^\pm - y^\pm-i\,\epsilon )\,
 \Big \}\,.
\ee

\no The two-point function for the scalar thermofield $
\phi (x ; \beta ) = \phi (x^+ ; \beta ) + \phi (x^- ; \beta )\,$ is given by,

\bear\label{2pfphi}
\langle 0, \widetilde 0 \vert \phi (x ; \beta ) \phi (y ; \beta ) 
\vert \widetilde 0, 0 \rangle &=&\frac{1}{\pi}
\mathit{z} (\beta , {\mu^\prime} )\\
&-&\frac{1}{4 \pi} \ln \Bigg \{-\mu^2\,\Big (\frac{\beta}{\pi} \Big )^2
\sinh \Big [ \frac{\pi (x^+ - y^+-i\,\epsilon )}{\beta } \Big ]
\sinh \Big [ \frac{\pi (x^- - y^--i\,\epsilon )}{\beta } \Big ] \Bigg \}.\nonumber
\ear

\no In the zero-temperature limit we obtain the two-point function of the
free massless scalar field,

\be
\langle 0, \widetilde 0 \vert \phi (x ; \beta ) \phi (y ; \beta ) 
\vert \widetilde 0, 0 \rangle_{T \rightarrow 0} \,\rightarrow\,\, \langle 0 \vert \phi (x) 
\phi (y) \vert 0 \rangle_o\,=\,-\,\frac{1}{4\pi}\,\ln\,\Big \{\,- \mu^2\,
(x - y)^2\,+\,i \epsilon (x_0 - y_0) \Big \}\,.
\ee

The fact that the short-distance behavior of the thermal two-point function is the same as
that for zero temperature is a crucial factor for the success of the 
thermofield bosonization scheme. 
The appearance of the additive temperature dependent arbitrary constant in
(\ref{btpf}), $\mathit{z}$, is of no immediate significance 
since the correlators of the massless scalar field in two dimensions
does not satisfy positivity. As we shall show in the following
section the dependency on this constant of the Wick ordered exponentials 
will disappear upon
imposing suitable superselection rules.

The off-diagonal contribution for the two-point function is given 
by \footnote{Gradshtein pag.494, eq. 18.},

$$
\langle 0, \widetilde 0 \vert \phi (x^\pm ; \beta ) 
\widetilde \phi (y^\pm ; \beta ) \vert \widetilde 0, 0 \rangle =\,
-\,\frac{1}{2 \pi}\,\int_{0}^{\infty}\,
\cos  p (x^\pm - y^\pm)\,e^{\frac{\beta\,p}{2}}\,N_B (\theta, p)\,
\frac{dp}{p}\,=
$$

$$-\,\frac{1}{4 \pi}\,\int_0^\infty\,
\frac{\cos p (x^\pm - y^\pm)}{\sinh \frac{\beta}{2}p}\,\frac{dp}{p}\,
$$

\be
= \, \frac{1}{4 \pi}\,\ln \,
\cosh \frac{\pi (x^\pm - y^\pm)}{\beta} \,-\,\frac{1}{4 \pi}\, 
f (\beta , \kappa ) \,,
\ee

\no where $f (\beta , \kappa )$ is the  integral,

\be
f (\beta , \kappa ) = \int_\kappa^\infty\,
\frac{1}{\sinh \frac{\beta}{2} p}\,\frac{d p}{p}\,
=\,\int_\kappa^\infty\,\frac{d p}{p}\,N_B (p; \beta )\,e^{\,\frac{\beta\,p}{2}}\,.
\ee

\no where $\kappa$ is a new  infrared cutoff. This integral
also play an important role in the selection rule of the off-diagonal
correlation functions of thermal Wick exponentials. In the 
limit $T \rightarrow 0$ or $x \approx y$, we get for the 
space-time dependent contribution,

\be
\langle 0, \widetilde 0 \vert \phi (x^\pm ; \beta ) 
\widetilde \phi (y^\pm ; \beta ) \vert \widetilde 0, 0 \rangle_{T \rightarrow 0} \rightarrow 0\,,
\ee

\no as expected.

\subsection{Wick Ordered Exponential and Selection Rules}

Let us consider the Wick ordered exponential of the 
free massless fields $\phi (x^\pm)$ and $\widetilde \phi(x^\pm)$,

\be
W (x ; \lambda ) = :e^{\,i\,\lambda\, \phi (x)}: \doteq e^{\,i\,\lambda\,\phi^{(-)}(x)}\,
e^{\,i\,\lambda\,\phi^{(+)}(x)}\,,
\ee

\be
\widetilde W (x ; \lambda ) = :e^{\,-\,i\,\lambda\, \widetilde \phi (x)}: \doteq
e^{\,-\,i\,\lambda\,\widetilde \phi^{(-)}(x)}\,
e^{\,-\,i\,\lambda\,\widetilde \phi^{(+)}(x)}\,.
\ee

\no The Wick exponential has to be understood as a formal series of
Wick-ordered powers of the field at the exponent such that,

\be
\langle 0 \vert W (x ; \lambda ) \vert 0 \rangle =
\langle \widetilde 0 \vert \widetilde W (x ; \lambda ) \vert \widetilde 0 \rangle = 1\,.
\ee

The unitary operator $U_B (\theta_B )$ acts on the creation and annihilation
components as

\be
U_B (\theta_B )\,\phi^{(\pm)}(x)\,U_B^{-1}(\theta_B ) = \phi_c^{(\pm)} (x; \beta ) - 
\widetilde \phi_s^{(\mp)} (x; \beta )\,,
\ee

\be
U_B (\theta_B )\,\widetilde \phi^{(+)}(x)\,U_B^{-1}(\theta_B ) = 
\widetilde \phi_c^{(\pm)} (x; \beta ) - 
\phi_s^{(\mp)} (x; \beta )\,.
\ee
 
\no The thermofield Wick exponentials are then given by,

$$
W (x ; \beta , \lambda ) = U_B ( - \theta_B ) W (x ; \lambda ) U_B ( \theta_B ) = 
\Big ( e^{\,i\,\lambda\,\phi^{(-)}_c (x; \beta )     \,-\,i\,\lambda\,\widetilde \phi^{(+)}_s (x; \beta )   }\,\Big )
\Big (e^{\,i\,\lambda\,\phi^{(+)}_c (x; \beta)\,
-\,i\,\lambda\,\widetilde \phi^{(-)}_s (x; \beta)}\,\Big )
$$

$$
= {\cal Z} (\beta , \mu^\prime , \lambda^2 )\,
\Big ( e^{\,i\,\lambda\,\phi^{(-)}_c (x; \beta )
\,-\,i\,\lambda\,\widetilde \phi^{(-)}_s (x; \beta)}\,\Big ) \Big (
e^{\,i\,\lambda\,\phi^{(+)}_c (x; \beta)\,
-\,i\,\lambda\,\widetilde \phi^{(+)}_s (x; \beta )} \Big )\,,
$$

$$
\widetilde W (x ; \beta , \lambda ) = U_B ( - \theta_B ) \widetilde W (x ; \lambda ) U_B ( \theta_B ) = 
\Big (e^{\,-\,i\,\lambda\,\widetilde \phi^{(-)}_c (x; \beta )+\,i\,\lambda\,\phi^{(+)}_s (x; \beta )}\Big )\,\Big (
e^{\,-\,i\,\lambda\,\widetilde \phi^{(+)}_c (x; \beta)+\,i\,\lambda\,\phi^{(-)}_s (x; \beta)}\Big )
$$

$$
= {\cal Z} (\beta , \mu^\prime, \lambda^2 )\,\Big (\,
e^{\,-\,i\,\lambda\,\widetilde \phi^{(-)}_c (x; \beta )
\,+\,i\,\lambda\,\phi^{(-)}_s (x; \beta)} \Big )\,\Big (
e^{\,-\,i\,\lambda\,\widetilde \phi^{(+)}_c (x; \beta)\,
+\,i\,\lambda\,\phi^{(+)}_s (x; \beta )} \Big )\,,
$$
 
\no where

\be
{\cal Z} ( \beta , \mu^\prime , \lambda^2 ) = e^{\,-\,\lambda^2\,
[ \widetilde \phi^{(+)}_s (x + \varepsilon; \beta )\,,
\,\widetilde \phi^{(-)}_s (x; \beta ) ]}\,
\Big \vert_{\varepsilon \rightarrow 0}\,=\,
e^{\,-\,\frac{\lambda^2}{4 \pi}\,
\int_{\mu^\prime}^\infty\,\frac{dk}{k}\,N_B (k; \beta )}\,,
\ee

\no that is

\be
{\cal Z}( \beta , \mu^\prime , \lambda^2)\,=\,e^{\,-\,\frac{\lambda^2}{4 \pi}\,\mathit{z}(\beta , \mu^\prime )}\,,
\ee

\no where $\mathit{z} (\beta , \mu^\prime )$ is the integral 
appearing in the two point
function (\ref{int}). The thermal Wick exponential in then given by,

\be\label{w1}
W (x ; \beta , \lambda ) = {\cal Z} (\beta , \mu^\prime , \lambda^2 )\,:e^{\,i\,\lambda\,\phi (x ; \beta )}:\,,
\ee

\be\label{w2}
\widetilde W (x ; \beta, \lambda ) = {\cal Z} (\beta , \mu^\prime , \lambda^2)
\,
:e^{\,-\,i\,\lambda\,\widetilde \phi (x ; \beta )}:\,.
\ee

\no Thus, we obtain,

\be
\langle 0(\beta ) \,\vert W (x;\lambda )\,\vert \,0(\beta ) \rangle\, = 
\,
\langle 0 , \widetilde 0\,\vert W (x ; \beta , \lambda )\,\vert \,\widetilde 0 , 0 \rangle\,
=\,{\cal Z} ( \beta , \mu^\prime , \lambda^2 )\,,
\ee

\be
\langle 0(\beta ) \,\vert \widetilde W (x ; \lambda )\,\vert \,0(\beta )  \rangle\, = 
\,
\langle 0 , \widetilde 0\,\vert \widetilde W (x ; \beta , \lambda )\,\vert \,\widetilde 0 , 0 \rangle\,
=\,{\cal Z} ( \beta , \mu^\prime , \lambda^2)\,.
\ee

\no As we shall see, the factor ${\cal Z} (\beta , \mu^\prime , \lambda^2 )$  
plays a role of a wave function 
renormalization for the bosonized Fermi thermofield. The exponential is IR 
dependent on the parameters $\mu$ and $\mu^\prime$, such that

\be
N_{\mu,\mu^\prime} \Big [e^{\,i\,\lambda\,\phi (x ; \beta)} \Big ]\,=\,
\Bigg (\frac{\mu^2}{\nu^2} \Bigg )^{\frac{\lambda^2}{4 \pi}}\,
\frac{\mathsf{Z} (\beta ; \mu^\prime)}{\mathsf{Z} (\beta ; \nu^\prime)}
\,N_{\nu, \nu^\prime} \Big [e^{\,i\,\lambda\,\phi (x ; \beta)} \Big ]\,,
\ee

\no where $N_{\mu,\mu^\prime}$ ($N_{\nu, \nu^\prime}$) stands for ``normal ordering'' 
with respect to the masses $\mu,\mu^\prime$ ($\nu,\nu^\prime$).

In order to display  the role played by the cut-off dependent 
term ${\cal Z} (\beta , \mu^\prime, \lambda^2 )$, let us 
consider the  correlation functions at 
finite temperature of the Wick exponential,

\be
W (x_j ; \lambda_j)\,=\,:\,e^{\,i\,\lambda_j\,\phi (x_j)}\,:\,.
\ee

\no  The diagonal contribution is given by,

$$
\langle 0(\beta )  \vert \prod_{j = 1}^n\,\,W (x_j ; \lambda_j )
\, \vert \,0(\beta ) \rangle \,=\,
\langle 0, \widetilde 0 \vert \prod_{j = 1}^n\, W (x_j; \beta , \lambda_j )\,
 \vert \widetilde 0, 0 
\rangle =
$$

$$
e^{\textstyle\,-\,\frac{1}{4 \pi} \,\mathit{z} (\beta , \mu^\prime )\sum_{j = 1}^n\,\lambda_j^2}\, \langle 0, \widetilde 0 \vert 
\prod_{j = 1}^n\,: e^{\,i\,\lambda_j\,
\phi_\beta (x_j) } :\, \vert \widetilde 0, 0 \rangle =
$$

$$
e^{\textstyle\,-\,\frac{1}{4 \pi} \,\mathit{z} (\beta , \mu^\prime )\,\sum_{j = 1}^n\,\lambda_j^2}\,
\prod_{i < j}^n\, e^{\,-\,\lambda_i\,\lambda_j\,
\{\,[ \phi^{(+)}_c (x_i; \beta ) , \phi^{(-)}_c (x_j; \beta )]
\,+\,
[\widetilde  \phi^{(+)}_s (x_i; \beta ) , 
\widetilde \phi^{(-)}_s (x_j; \beta )]\,\}}\,=
$$

\be
e^{\textstyle\,-\,\frac{1}{4 \pi}\,\mathit{z} (\beta , \mu^\prime )\,\sum_{j=1}^n \lambda_j^2}\,
\prod_{i < j}^n\, e^{\,-\,\lambda_i\,\lambda_j\,\{ 
D_c^{(+)} (x_i - x_j; \beta ) + \widetilde D_s^{(+)} (x_i - x_j; \beta )\}}\,.
\ee

\no Using the expression (\ref{btpf}) for the thermal two-point function, we obtain,

\begin{eqnarray}\label{selection}
\langle 0(\beta )  \vert \prod_{j = 1}^n\,\,W (x_j ; \lambda_j)
\, \vert\, 0(\beta )  
\rangle \,=\,&{}&
e^{\textstyle\,-\,\frac{1}{4 \pi}\,
\mathit{z} (\beta , \mu^\prime )\Big ( \sum_{j = 1}^n\,\lambda_j \Big )^2 }\,
\times\nonumber\\
&&\;\;\;\;\;\,
\prod_{i < j}^n\,\Big [ i \mu \,\frac{\beta}{\pi}\,
\sinh \frac{\pi (x_i - x_j-i\,\epsilon )}{\beta}\,
\Big ]^{\,\frac{\lambda_i \lambda_j}{4 \pi}}\,.
\end{eqnarray}

\no In order to ensure that 
the correlation functions of the 
Wick exponential at finite temperature be
independent of $\mu^\prime$, we must impose 
the selection rule,

\be\label{sr}
\sum_{i = 1}^n \lambda_i\,=\,0\,.
\ee

\no Thus, we can write,

$$
\langle 0(\beta ) \, \vert \prod_{j = 1}^n\,\,W (x_j ; \lambda_j)
\, \vert \,0(\beta )  
\rangle \,=\,{\LARGE \delta}_{_{\!\Sigma \lambda_i , 0}}\,\,
\langle 0 , \widetilde 0 \vert\,\prod_{j = 1}^n\,W (x_j ; \beta , \lambda_j )\,\vert
\widetilde 0 , 0 \rangle\,=
$$

$$
=\,{\LARGE \delta}_{_{\Sigma \lambda_i , 0}}\,\,\Big (\,\mu\,\Big )^{\textstyle
\,\sum_{i < j}^n \frac{\lambda_i\,\lambda_j}{4 \pi}}\,
\prod_{i < j}^n\Big [ \,\frac{\beta}{\pi}\,
\sinh \frac{\pi (x_i - x_j-i\,\epsilon )}{\beta}\,
\Big ]^{\textstyle\,\frac{\lambda_i \lambda_j}{4 \pi}}\,
$$

\be\label{cfwe}
\equiv\,{\LARGE \delta}_{_{\Sigma \lambda_i , 0}}\,\,
\prod_{j = 1}^{n}\,\Big ( \mu\,\Big )^{\textstyle\,
-\,\frac{\lambda_j^2}{8 \pi}}\,
\prod_{i < j}^n\Big [ \,\frac{\beta}{\pi}\,
\sinh \frac{\pi (x_i - x_j-i\,\epsilon )}{\beta}\,
\Big ]^{\textstyle\,\frac{\lambda_i \lambda_j}{4 \pi}}\,
\ee

\no As in the $T = 0$ case we may thus define a multiplicatively, infrared
renormalized Wick exponential,

\be
W_R (x_j ; \lambda_j)\, =\, \Big (\,\mu\, \Big )^{\frac{\lambda^2_j}{8 \pi}}\,
W (x_j ; \lambda_j)\,,
\ee

\no such that,

\be\label{rwe}
W_R (x_j ; \beta , \lambda_j )\,=\,\Big (\,\mu \,
\Big )^{\frac{\lambda^2_j}{8 \pi}}\,
{\cal Z} (\beta, \mu^\prime , \lambda_j^2 )\,
:e^{\,i\,\lambda_j\,\phi (x_j ; \beta )}:\,.
\ee

\no It is interesting that one and the same selection 
rule $\sum_i \lambda_i = 0$ allows us to eliminate the 
dependence on the infrared regulators $\mu$ and $\mu^\prime$. Thus, provided 
we associate with  the Wick exponential a conserved charge $\lambda$, the 
correlation functions of the thermal Wick exponentials (\ref{rwe}) 
are mapped into Wightman functions  belonging to a positive metric
Hilbert space. In a similar way for the correlation functions of 
the Wick exponential $\widetilde W (x_j ; \beta, \lambda_j )$,

\be\label{rtwe}
\widetilde W_R (x_j ; \beta , \tilde\lambda_j )\,=\,\Big (\,\mu \,
\Big )^{\frac{\tilde\lambda^2_j}{8 \pi}}\,
{\cal Z} (\beta , \mu^\prime , \tilde\lambda_j^2 )\,:e^{\,-\,i\,
\tilde\lambda_j\,
\widetilde \phi (x_j ; \beta )}:\,.
\ee
 
\no Using Eq. (\ref{btpf}), the correlation functions of the Wick
exponential $\widetilde W (x ; \beta , \lambda )$ can be obtained 
from (\ref{cfwe}) by making  $i \epsilon \rightarrow - i \epsilon$ and
introducing a phase factor

\be\label{phase}
e^{\,i\,\pi\,\sum_{i < j}^n\, \frac{\lambda_i \lambda_j}{4 \pi}}\,.
\ee

The off-diagonal correlation functions are given by,

$$
\langle 0(\beta )  \vert \prod_{i = 1}^n\,W (x_i ; \lambda_j)\,
\prod_{j = 1}^m\,\widetilde W (y_j ; \tilde\lambda_j) \vert\, 0(\beta )  \rangle \,=
\mu^{\, \frac{1}{8\pi}\,\Big \{\,\Big ( \sum_i^n\,\lambda_i \Big )^2\,
+\,\Big (  \sum_j^m \tilde\lambda_j \Big )^2 \Big \} }\,\times
$$

$$
e^{\,-\,\frac{1}{8 \pi}\,\mathit{z} (\beta , \mu^\prime )\,\Big \{\,
\Big ( \sum_{i}^n\,\lambda_i \Big )^2\,+\,\Big ( \sum_{j}^m\,
\tilde\lambda_j \Big )^2\,
\Big \}}\,e^{\textstyle\,-\,\frac{1}{8 \pi}
\,f (\beta , \kappa)\,\Big ( \sum_i^n\,\lambda_i\, \Big )\,
\Big ( \sum_j^m\,\tilde\lambda_j \Big )}\,\times
$$

$$
\prod_{i < k}^n\,\Big [ \,\frac{\beta}{\pi}\,
\sinh \frac{\pi (x_i - x_k)}{\beta}\,
\Big ]^{\textstyle\,\frac{\lambda_i \lambda_k}{4 \pi}}\,
\prod_{j < \ell}^m\,\Big [ \,\frac{\beta}{\pi}\,
\sinh \frac{\pi (y_j - y_\ell)}{\beta}\,
\Big ]^{\textstyle\,\frac{\tilde\lambda_j \tilde\lambda_\ell}{4 \pi}}\,\times
$$

\be
\prod_{i , j}^{n,m}\,\Big [ \,\cosh \frac{\pi (x_i - y_j)}{\beta}\,
\Big ]^{\textstyle\,-\,\frac{\lambda_i \tilde\lambda_j}{4 \pi}}\,.
\ee

\no Again, the selection rules that makes the off-diagonal correlation 
function simultaneously 
independent of $\mu$, $\mu^\prime$ and  $\kappa $ are,

\be
\sum_{i = 1}^n\,\lambda_i\,=\,0\,\,\,,\,\,\,
\sum_{j = 1}^m\,\tilde\lambda_j\,=\,0\,,
\ee

and we get for the IR renormalized Wick exponentials,

$$
\langle 0(\beta )  \vert \prod_{i = 1}^n\,W (x_i; \lambda_i)\,
\prod_{j = 1}^m\,\widetilde W (y_j ; \tilde\lambda_j) \vert \,0(\beta )  \rangle \,=
{\LARGE \delta}_{_{\!\Sigma \lambda_i , 0}}\,\,
{\LARGE \delta}_{_{\!\Sigma \tilde\lambda_j , 0}}\,\,
\times
$$

$$
\prod_{i < k}^n\,\Big [ \,i\,\frac{\beta}{\pi}\,
\sinh \frac{\pi (x_i - x_k)}{\beta}\,
\Big ]^{\textstyle\,\frac{\lambda_i \lambda_k}{4 \pi}}\,
\prod_{j < \ell}^m\,\Big [ \,i\,\frac{\beta}{\pi}\,
\sinh \frac{\pi (y_j - y_\ell)}{\beta}\,
\Big ]^{\textstyle\,\frac{\tilde\lambda_j \tilde\lambda_\ell}{4\pi}}\,\times
$$

\be
\prod_{i , j}^{n,m}\,\Big [ \,\cosh \frac{\pi (x_i - y_j)}{\beta}\,
\Big ]^{\textstyle\,-\,\frac{\lambda_i \tilde\lambda_j}{4\pi}}\,.
\ee

\section{Free Massless Fermi Thermofield}
\setcounter{equation}{0}

Let us introduce the fermion doublet \cite{Ojima}

\be\label{fd}
\Psi  = \pmatrix{\psi  \cr \phantom{..........} \cr i\,\widetilde \psi^\ast }\,,
\ee

\no and the corresponding total Lagrangian density,

\be
{\cal L}_T  = i\,\overline\Psi  \gamma^\mu \partial_\mu \Psi\,
= \,{\cal L} - \widetilde{\cal L}\,=\,
i\,\overline{\psi}  \gamma_\mu \partial_\mu \psi\,-\,\Big ( - i\, \overline{\widetilde\psi} \gamma^\mu \partial_\mu
\widetilde\psi \Big )\,.
\ee

\no The free massless Fermi fields $\psi$ and $\widetilde\psi$ are given 
in terms of right and left spinor components,

\be
\psi (x) = \pmatrix{\psi (x^+) \cr \psi (x^-)}\,,
\ee

\be
\widetilde\psi (x) = \pmatrix{\widetilde\psi (x^+) \cr \widetilde\psi (x^-)}\,,
\ee

\no where

\be
\psi (x^\pm)\,=\,\int_0^\infty\,\frac{dp}{\sqrt{2 \pi}}\,\big \{
\,f_p (x^\pm)\,b (\mp p)\,+\,f^\ast_p (x^\pm)\,
d^\dagger (\mp p)\, \big \}\,,
\ee

\be
\widetilde \psi (x^\pm)\,=\,
\int_0^\infty\,\frac{dp}{\sqrt{2 \pi}}\,\big \{
\,f^\ast_p (x^\pm)\,\widetilde b (\mp p)\,+\,
f_p (x^\pm) \big \}\,\widetilde d^\dagger (\mp p)\,\,.
\ee

The unitary operator taking one to the thermofields is now given by,

\be
U_F (\theta_F )\,=\,e^{\,-\,\int_{-\infty}^\infty \,d\,p\,\theta_F (\vert p^1\vert  ,\beta )\Big ( \widetilde b (p^1)\,
b (p^1)\,-\,b^\dagger (p^1)\,\widetilde{b}^\dagger (p^1)\,+\, 
\widetilde d (p^1)\,
d (p)\,-\,d^\dagger (p^1)\,\widetilde{d}^\dagger (p^1)\,\Big )}\,,
\ee

\no and the corresponding transformed annihilation operators are

\be
b (p^1 ; \beta )\,=\,b (p^1)\, \cos \theta_F (p ; \beta )\,-\,
\widetilde{b}^\dagger (p^1)\,\sin \theta_F (p ; \beta )\,,
\ee

\be
\widetilde b (p^1 ,\beta )\,=\,\widetilde b (p^1)\, \cos_F \theta (p ; \beta )\,+\,
{b}^\dagger (p^1)\,\sin \theta_F (p ; \beta )\,,
\ee

\no with similar expressions for $d (p^1 ; \beta )$ 
and $\widetilde d (p^1 ; \beta )$, where the Bogoliubov 
parameter $\theta_F (p , \beta )$ is now
implicitly defined by,

\be
\cos \theta_F (p ; \beta )\,=\,
\frac{1}{\sqrt{1 + e^{\,-\,\beta\,p}}}\,,
\ee

\be
\sin \theta_F (p ; \beta )\,=\,
\frac{e^{\,-\,\beta\,p /2}}{\sqrt{1 + e^{\,-\,\beta\,p}}}\,,
\ee

\no and the Fermi-Dirac statistical  weight is given by,

\be
N_F (p; \beta )\, = \,\sin^2 \theta_F (p ; \beta )\,=\,\frac{1}{e^{\,\beta p} + 1}\,.
\ee

\no The  fermion thermofields are then

$$
\psi (x^\pm ; \beta ) = 
\int_0^\infty\,\frac{dp}{\sqrt{2 \pi}}\,\Big \{
f_p (x^\pm) \Big ( b (\mp p)\,\cos \theta_F (p ; \beta ) - 
\widetilde b^\dagger (\mp p)\sin \theta_F (p ; \beta ) \Big )
$$

\be   +\,
 f^\ast_p (x^\pm) \Big ( d^\dagger (\mp p)\,\cos \theta_F (p ; \beta ) - 
 \widetilde d (\mp p)
\sin \theta_F (p ; \beta ) \Big )\,\Big \}\,.
\ee

\no and

$$
\widetilde \psi (x^\pm ; \beta ) = 
\int_0^\infty\,\frac{dp}{\sqrt{2 \pi}}\,\Big \{
f^\ast_p (x^\pm) \Big (\widetilde b (\mp p)\,\cos \theta_F (p ; \beta ) +
b^\dagger (\mp p)\sin \theta_F (p ; \beta ) \Big )
$$

\be   +\,
f_p (x^\pm) \Big ( \widetilde d^\dagger (\mp p)\,\cos \theta_F (p ; \beta ) +  
d (\mp p)\sin \theta_F (p ; \beta ) \Big )\,\Big \}\,.
\ee

\no Besides the canonical anticommutation relations we have,

\be
\{ \psi (x ; \beta )\,,\,\widetilde \psi (y ; \beta ) \}\,=\,0\,\,\,,\,\,\,
\{ \psi^\dagger (x ; \beta )\,,\,\widetilde \psi (y ; \beta ) \}\,=\,0\,.
\ee

\subsection{Fermion Thermofield Two-point Function}

The diagonal contribution for the two-point function of the thermofield fermion
is given by,

\be\label{tpf}
\langle 0, \widetilde 0 \vert \psi (x^\pm ; \beta ) 
\psi^\dagger (y^\pm ; \beta ) \vert 0 , \widetilde 0
\rangle = \,\frac{1}{2\, i \,\pi (x^\pm - y^\pm)}\,-\,\frac{1}{i\,\pi}\,
\int_0^\infty\,d p\,N_F (p; \beta )\,\sin p (x^\pm - y^\pm)\,.
\ee

\no The first term in (\ref{tpf}) is the two-point function at zero temperature,

\be
\langle 0 \vert \psi^\dagger (x^\pm) \psi (y^\pm) \vert 0 \rangle_0 = \,
\frac{1}{2\, i \,\pi (x^\pm - y^\pm)}\,.
\ee

\no The integral appearing in the two-point function (\ref{tpf}) is evaluated to be
\footnote{Gradshtein - 4 Edition,
pag. 481- eq. 3.9911-1,

$$
\int_0^\infty\,dx\,\frac{\sin ax}{e^{\beta x} + 1} = \frac{1}{2 a}\,-\,
\frac{\pi}{2 \beta \sinh \frac{a \pi}{\beta}}\,.
$$
}

\be
\frac{1}{i\,\pi}\,
\int_0^\infty\,d p\,N_F (p; \beta )\,\sin p (x - y) =
\,\frac{1}{2\, i \,\pi (x - y) }\,-\,
\frac{1}{2 i \beta \sinh [ \frac{\pi}{\beta} (x - y)]}\,.
\ee

\no We obtain after appropriate $i\epsilon$ prescription,

\be\label{ftpf1}
\langle 0, \widetilde 0 \vert \psi (x^\pm ; \beta ) 
\psi^\dagger (y^\pm ; \beta ) \vert 0 , \widetilde 0
\rangle\, =\,
\frac{1}{2 i\,\beta\,\sinh [\frac{\pi}{\beta} (x^\pm - y^\pm-i\,\epsilon )]}\,.
\ee

\be\label{tftpf1}
\langle 0, \widetilde 0 \vert \widetilde{\psi} (x^\pm ; \beta ) 
\widetilde \psi^\dagger (y^\pm ; \beta ) \vert 0 , \widetilde 0
\rangle\, =\,
\frac{-\,1}{2 i\,\beta\,\sinh [ \frac{\pi}{\beta} (x^\pm - y^\pm+i\,\epsilon )]}\,.
\ee

The diagonal thermofield fermion correlation function satisfies the asymptotic 
factorization property

\be
\lim_{\lambda \rightarrow \infty}\,
\langle 0, \widetilde 0 \vert \psi (x + \lambda ; \beta ) 
\psi^\dagger (x ; \beta ) \vert 0 , \widetilde 0
\rangle\,\rightarrow\,\Big \vert \langle 0, \widetilde 0 \vert 
\psi (x ; \beta ) \vert 0 , \widetilde 0
\rangle \Big \vert^2\,=\,0\,.
\ee

For $T \rightarrow 0$ as well as $x \approx y$ at fixed $T$, we obtain,

\be
\langle 0, \widetilde 0 \vert \psi (x^\pm ; \beta ) \psi^\dagger (y^\pm ; \beta ) \vert 0 , \widetilde 0
\rangle\, \rightarrow\,
\frac{1}{2 i\,\pi (x^\pm - y^\pm - i\,\epsilon)}\,.
\ee

For the off-diagonal contribution we find \footnote{Gradshtein, pag. 503, last integral.},

\bear
\langle 0, \widetilde 0 \vert \widetilde \psi (x^\pm ; \beta ) 
\psi (y^\pm ; \beta ) \vert 0 , \widetilde 0
\rangle \,&=&\,
-\,\frac{1}{ \pi}\,\int_0^\infty\,\cos p (x^\pm - y^\pm )\,N_F(\beta, p )\,e^{\,-\,\frac{\beta}{2} p}\,d p\,
\\ \nonumber
&=&
-\,\frac{1}{2 \pi}\,\int_0^\infty\,\frac{\cos p (x^\pm - y^\pm )}{\cos \frac{\beta}{2} p}\,d p \\
\nonumber
&=&\,  -\,\frac{1}{2 \beta \cosh [ \frac{\pi}{\beta}(x^\pm - y^\pm)]}\,.
\ear

\no In the limit $T \rightarrow 0$ we get,

\be
\langle 0, \widetilde 0 \vert \widetilde \psi (x^\pm ; \beta ) 
\psi (y^\pm ; \beta ) \vert 0 , \widetilde 0
\rangle_{T \rightarrow 0}\,\rightarrow\,0\,,
\ee
 
\no as expected. The off-diagonal two-point function also satisfy 
the cluster decomposition
property. Moreover,

\be\label{ncf}
\langle 0, \widetilde 0 \vert \widetilde \psi (x^\pm ; \beta ) 
\psi^\dagger (y^\pm ; \beta ) \vert 0 , \widetilde 0 \rangle = 0\,.
\ee

\section{Thermofield Bosonization of the Free Massless Fermion}
\setcounter{equation}{0}

The scale dimension of the thermofield Wick exponential is $d = 
\frac{\lambda^2}{4 \pi}$. In analogy to the $T = 0$ case, we seek 
a representation of the thermal Fermi field in terms of the IR renormalized
thermofield Wick exponential of canonical scale 
dimension $\lambda = 2 \sqrt \pi$:

\be\label{bfo}
\psi (x^\pm ; \beta ) \doteq \,W_R (x^\pm ; \beta ) = \,
C\,{\cal Z}_c (\beta , \mu^\prime )\,
:e^{\,2 i \sqrt \pi\,\phi (x^\pm ; \beta )}:\,,
\ee
\be\label{bfo2}
\widetilde\psi (x^\pm ; \beta ) \doteq \,\widetilde W_R (x^\pm; \beta ) =\,
 C\,{\cal Z}_c (\beta , \mu^\prime )\,
: e^{\,-\,2 i \sqrt \pi\,\widetilde\phi (x^\pm ; \beta )}:\,.
\ee
where
\be
W_R(x^\pm ;\beta)= W_R(x^\pm; \beta,2\sqrt{\pi})\nonumber
\ee
\be
{\cal Z}_c(\beta,\mu^\prime) = {\cal Z} (\beta,\mu^\prime,4\pi)\nonumber
\ee 
and 
\be
C = \Big ( \frac{\mu}{2 \pi} \Big )^{1/2}\,.
\ee

Taking into account the super selection rule associated 
with the Wick exponential of thermofields, one readily verifies that the set of all 
two-point functions  of the 
Fermi thermofield, obtained in the previous section, are 
recovered from the bosonized expression (\ref{bfo}). In particular 
property (\ref{ncf}) follows from the superselection rule for 
the  Wick exponential of the Bose thermofield. The minus sign appearing
in the two-point function for the thermofield $\widetilde \psi (x ; \beta )$ 
arises from the phase factor (\ref{phase}).

\subsection{Statistics}

Let us consider the anticommutation relation between the same spinor 
components labeled by $\pm$. From (\ref{bfo}) we obtain,

\bear 
\psi (x^\pm ; \beta )\psi (y^\pm ; \beta ) &=& C^2\,{\cal Z}_c^2 (\beta , \mu^\prime )\,
:e^{\,2\,i\,\sqrt \pi\,\phi (x^\pm ; \beta )}:\,
:e^{\,2\,i\,\sqrt \pi\,\phi (y^\pm ; \beta )}: \\ \nonumber
&=&
C^2\,{\cal Z}_c^2 (\beta , \mu^\prime )\,:e^{\,2\,i\,\sqrt \pi \,
[ \phi (x^\pm; \beta ) + \phi (y^\pm; \beta ) ]}:
\,e^{\,-\,4 \pi\,\{D_c^{(+)} (x^\pm - y^\pm; \beta ) + 
\widetilde D_s^{(+)} ( x^\pm - y^\pm; \beta ) \}}\\ \nonumber
&=&
\,C^2\,{\cal Z}_c^4 ( \beta , \mu^\prime )\,
:\psi (x^\pm ; \beta )\psi (y^\pm ; \beta ):\,\,
\Big ( i\,\mu\,
\frac{\beta}{\pi}\,\sinh [ \frac{\pi}{\beta} (x^\pm - y^\pm -i\epsilon)]\,\Big )\,.
\ear

\no From the antisymmetry of the $\sinh z$  under $z\rightarrow -z$
it follows that,

\be
\{ \psi (x^\pm ; \beta ), \psi (y^\pm ; \beta ) \} \,= \,0\,.
\ee

\no In a similar way, we obtain,

$$
\{ \psi (x^\pm ; \beta )\,,\,\psi^\dagger (y^\pm ; \beta ) \}\, = 
\,:e^{\,2\,i\,\sqrt \pi \,
[ \phi (x^\pm; \beta ) - \phi (y^\pm; \beta ) ]}:\,\times
$$

\be\label{facr}
\frac{1}{2 i \beta}\,\Bigg (
\, \frac{1}{\sinh [\frac{\pi}{\beta} (x^\pm - y^\pm -i\,\epsilon )]}\,
-\,\frac{1}{\sinh [\frac{\pi}{\beta} (x^\pm - y^\pm + i \epsilon )]}\Bigg )\,.
\ee
 
\no The non zero contributions in Eq. (\ref{facr}) arises from the neighborhood
of the point $x = y$. This time we get,

\be
\{\psi (x^\pm ; \beta )\,,\,\psi^\dagger (y^\pm ; \beta )\}\,
=\,\delta ( x^\pm - y^\pm )\,.
\ee

For the tilde-fields we analogously find,

\be
\{ \widetilde\psi (x^\pm ; \beta ), \widetilde\psi (y^\pm ; \beta ) \} \,= \,0\,,
\ee

\be
\{ \widetilde\psi (x^\pm ; \beta )\,,\,\widetilde\psi^\dagger (y^\pm ; \beta )\}\,
=\,\delta ( x^\pm - y^\pm )\,.
\ee

In order to ensure the correct anticommutation relations between 
$\psi (x^+ ; \beta )$ and $\psi (x^- ; \beta )$, between 
$\widetilde\psi (x^+ ; \beta )$ and $\widetilde\psi (x^- ; \beta )$, as
well as the anticummutativity of tilde operators with non-tilde ones, Klein 
factors \cite{Klein,Jost} must be introduced. To begin with, let
us introduce the charges,

\be
{\cal Q}_\beta^\pm\,\doteq\,\int_{- \infty}^{+ \infty}\,
\partial_0\, {\phi} (x^\pm ; \beta )\,d x^1\,,
\ee

\be
\widetilde{\cal Q}_\beta^\pm\,\doteq\,\int_{- \infty}^{+ \infty}\,
\partial_0\, {\widetilde\phi} (x^\pm ; \beta )\,d x^1\,,
\ee

\no and the Klein factors,

\be
K_\beta^\pm\,=\,e^{\,i\,\frac{\sqrt \pi}{2}\,{\cal Q}^\pm_{_{\!\beta}}}\,,
\ee

\be
\widetilde K_\beta^\pm\,=\,e^{\,-\,i\,\frac{\sqrt \pi}{2}\,
\widetilde{\cal Q}^\pm_{_{\!\beta}}}\,.
\ee

\no Since

\be
{\cal Q}_\beta^\pm = U_B ( - \theta ) {\cal Q}^\pm U_B (\theta )\,,
\ee

\no the Klein factor $K_\beta^\pm$ is the transformed Klein factor of the
Fermi field at zero temperature,

\be
K_\beta^\pm = U_B ( - \theta ) K^\pm U_B (\theta )\,.
\ee

\no Starting from eqs. (\ref{bfo}) and (\ref{bfo2}) the thermofields are redefined as,

\be\label{k1}
\psi (x^+ ; \beta )\,=\,W_R (x^+ ; \beta )\,,
\ee

\be\label{k2}
\psi (x^- ; \beta )\,=\,K^+_\beta\,W_R (x^- ; \beta )\,,
\ee

\be\label{k3}
\widetilde\psi (x^+ ; \beta )\,=\,K^+_\beta\,K^-_\beta\,
\widetilde W_R (x^+ ; \beta )\,,
\ee

\be\label{k4}
\widetilde\psi (x^- ; \beta )\,=\,K^+_\beta\,K^-_\beta\,
\widetilde K^+_\beta\,
\widetilde{W}_R (x^- ; \beta )\,.
\ee

\no These fields obey the correct anticommutation relations. The 
factor $K^+_\beta$ in (\ref{k2}) and $\widetilde K^+_\beta$ in (\ref{k4})
ensures that for different (left , right ) spinor components we obtain,

\be
\{ \psi(x^\pm ; \beta )\,,\,\psi (y^\mp ; \beta ) \}\,=\,0\,,
\ee

\be
\{ \widetilde\psi (x^\pm ; \beta )\,,\,\widetilde\psi (y^\mp ; \beta ) \}\,=\,0\,.
\ee

\no The factors $K^\pm_\beta$ in (\ref{k3}) and (\ref{k4}) ensures the normal
anticommutativity between tilde thermofermion components and non-tilde
ones,

\be
\{ \widetilde\psi (x^\pm ; \beta )\,,\,\psi (y^\pm ; \beta ) \}\,=\,0\,,
\ee

\be
\{ \widetilde\psi (x^\pm ; \beta )\,,\,\psi (y^\mp ; \beta ) \}\,=\,0\,.
\ee

\subsection{Fermionic Current and $\varepsilon$-expansion }

The vector current corresponding to the fermion doublet (\ref{fd}) is given by,

\be
\overline{\Psi} \gamma^\mu \Psi = {\cal J}^\mu - \widetilde {\cal J}^\mu\,,
\ee

\no where,

\be
{\cal J}^\mu = \overline{\psi} \gamma^\mu \psi\,,
\ee

\be
\widetilde {\cal J}^\mu =   \overline{\widetilde\psi} \gamma^\mu \widetilde\psi\,.
\ee

\no Let us define the finite-temperature left and right current 
components 

\be
J^\pm (x ; \beta ) \,=\,\psi^\dagger (x^\pm ; \beta )\,
\psi (x^\pm ; \beta )\,,
\ee

\be
\widetilde J^\pm (x ; \beta ) \,=\,\widetilde\psi^\dagger (x^\pm ; \beta )\,
\widetilde\psi (x^\pm ; \beta )\,,
\ee

\no by   point-splitting as,

\be
J^\pm (x ; \beta )\,=\,\lim_{\varepsilon^\pm \rightarrow 0}\,\Big \{\,
\psi^\dagger (x^\pm  + \varepsilon^\pm; \beta )\,\psi (x^\pm ; \beta )\,-\,
\langle 0, \widetilde 0 \vert\, \psi^\dagger (x^\pm + \varepsilon^\pm; \beta )\,
\psi (x^\pm ; \beta )\,
\vert 0, \widetilde 0 \rangle\,\Big \}\,,
\ee
 
\no and similarly for $\widetilde J^\pm (x ; \beta )$. We obtain,

$$
J^\pm (x ; \beta ) = \frac{\mu}{2 \pi}\,{\cal Z}_c^2 (\beta , \mu^\prime )\,
\lim_{\varepsilon^\pm \rightarrow 0}\,\,\Big \{\,
: e^{\,-\,2\,i\,\sqrt \pi\,\phi (x^\pm + \varepsilon^\pm ; \beta )} :\,
: e^{\,2\,i\,\sqrt \pi\,\phi (x^\pm ; \beta )}:\,-\,(V.E.V.)_{\beta}\,
\Big \}\,=\,
$$

\be
\frac{\mu}{2 \pi}\,{\cal Z}_c^2 (\beta , \mu^\prime )\,
\lim_{\varepsilon^\pm \rightarrow 0}\,\,\Big \{\,
: e^{\,-\,2\,i\,\sqrt \pi\,[\phi (x^\pm + \varepsilon^\pm ; \beta )\,
-\,\phi (x^\pm ; \beta )]} :\,e^{\,4\,\pi\,D^{(+)}_c (\varepsilon^\pm; \beta )
\,+\,4\,\pi\,\widetilde D^{(+)}_s ( \varepsilon^\pm; \beta )}\,-\,(V.E.V.)_{\beta}\,
\Big \}\,,
\ee

$$
\widetilde J^\pm (x ; \beta ) = \frac{\mu}{2 \pi}\,{\cal Z}_c^2 (\beta , \mu^\prime )\,
\lim_{\varepsilon^\pm \rightarrow 0}\,\,\Big \{\,
: e^{\,2\,i\,\sqrt \pi\,\widetilde\phi (x^\pm + \varepsilon^\pm ; \beta )} :\,
: e^{\,-\,2\,i\,\sqrt \pi\,\widetilde\phi (x^\pm ; \beta )}:\,-\,(V.E.V.)_{\beta}\,
\Big \}\,=\,
$$

\be
\frac{\mu}{2 \pi}\,{\cal Z}_c^2 (\beta , \mu^\prime )\,
\lim_{\varepsilon^\pm \rightarrow 0}\,\,\Big \{\,
: e^{\,2\,i\,\sqrt \pi\,[\widetilde\phi (x^\pm + \varepsilon^\pm ; \beta )\,
-\,\widetilde\phi (x^\pm ; \beta )]} :\,e^{\,4\,\pi\,\widetilde D^{(+)}_c (\varepsilon^\pm; \beta )
\,+\,4\,\pi\, D^{(+)}_s ( \varepsilon^\pm; \beta )}\,-\,(V.E.V.)_{\beta}\,
\Big \}\,.
\ee
 
\no Since the short-distance behavior of
the thermal two-point functions (\ref{btpf}) and (\ref{tbtpf}) is the same
as that for zero temperature, and the superselection
rule is identically satisfied, we obtain 
the $\mu$ and $\mu^\prime$ independent result,

\begin{eqnarray}
J(x^\pm ; \beta ) &=& \lim_{\varepsilon^\pm \rightarrow 0}\,
\frac{1}{2 \,\pi\,i\,\varepsilon^\pm}\,\Big \{\,
: e^{\,-\,2\,i\,\sqrt \pi\,\varepsilon^\pm\, \partial_{x^\pm}\,
\phi (x^\pm ; \beta )}\,
:\,-\,(V.E.V.)_{\beta}\,\Big \}\,,\nonumber\\
&=& \,-\frac{1}{\sqrt \pi}\,\partial_{x^\pm}\,
\phi (x^\pm ; \beta )\,,\label{bc}
\end{eqnarray}

\begin{eqnarray}
\widetilde J(x^\pm ; \beta ) &=& \lim_{\varepsilon^\pm \rightarrow 0}\,
\frac{e^{\,-\,i\,\pi}}{2 \,\pi\,i\,\varepsilon^\pm}\,\Big \{\,
: e^{\,2\,i\,\sqrt \pi\,\varepsilon^\pm\, \partial_{x^\pm}\,
\widetilde\phi (x^\pm ; \beta )}\,
:\,-\,(V.E.V.)_{\beta}\,\Big \}\,,\nonumber\\
&=& \,-\frac{1}{\sqrt \pi}\,\partial_{x^\pm}\,
\widetilde\phi (x^\pm ; \beta )\,,\label{tbc}
\end{eqnarray}

\no Defining the pseudoscalar thermofield $\varphi (x ; \beta )$, such 
that,

\be
\epsilon_{\mu \nu} \partial^\nu \varphi (x ; \beta ) = \partial_\mu \phi (x ; \beta )\,
\ee

\no the vector currents can be written as,

\be
{\cal J}^\mu (x ; \beta )\,=\,:\overline\psi (x ; \beta ) \gamma^\mu \psi (x ; \beta ):\,=\,-\,\frac{1}{\sqrt \pi}\,\epsilon^{\mu \nu}\,
\partial_\nu\,\varphi (x ; \beta )\,=\,
-\,\frac{1}{\sqrt \pi}\,\partial^\mu\,\phi (x; \beta )\,,
\ee
 
\be
\widetilde{\cal J}^\mu (x ; \beta )\,=\,:\widetilde{\overline\psi} (x ; \beta ) \gamma^\mu \widetilde \psi (x ; \beta ):\,=\,-\,
\frac{1}{\sqrt \pi}\,\epsilon^{\mu \nu}\,
\partial_\nu\,\widetilde\varphi (x ; \beta )\,=\,
-\,\frac{1}{\sqrt \pi}\,\partial^\mu\,\widetilde\phi (x; \beta )\,.
\ee
\

From (\ref{bc}), the current two-point function is found to be given by,

\begin{eqnarray}
\langle 0, \widetilde 0 \vert J^\pm (x ; \beta ) J^\pm (y ; \beta ) \vert 0, 
\widetilde 0 \rangle\,=&{}&\,\frac{1}{\pi}\,
\,\partial_{x^\pm}\,\partial_{y^\pm}\,
\langle 0, \widetilde 0 \vert \phi (x ; \beta ) \phi (y ; \beta ) 
\vert 0, \widetilde 0 \rangle \nonumber \\
&&\;\;\;\;= \frac{-1}{4 \beta^2}\,\frac{1}{\sinh^2 
 [ \frac{\pi}{\beta} (x^\pm - y^\pm - i \epsilon ) ]}\,.
\end{eqnarray}

\no In the zero temperature limit we recover the zero temperature correlation
function,

\be
\langle 0 \vert J^\pm (x) J^\pm (y) \vert 0 \rangle\,=\,\frac{-1}{4 \pi^2}\,
\frac{1}{(x^\pm\,-\,y^\pm - i \epsilon)^2}\,.
\ee

An alternative way to obtain the bosonized form of the current (\ref{bc}) is 
to consider the expectation value of the $T = 0$ current with respect to 
the $\vert \beta , 0 \rangle$ vacuum. To this end, let us again define
the current
two-point function in terms of a point-splitting,  

$$
\langle 0 (\beta ) \vert J^\pm (x)\,J^\pm (y ) \vert 0 (\beta ) \rangle\,=
$$

$$
\lim_{\varepsilon^\pm ,\,\delta^\pm \,\rightarrow 0}\,\langle 0 (\beta ) \vert\,
\Big (\,\psi^\dagger (x^\pm + \varepsilon^\pm)\,
\psi (x^\pm)\,\Big )\,\Big (\,\psi^\dagger (y^\pm + \delta^\pm )\,
\psi (y^\pm)\,\Big )\,\vert 0 (\beta ) \rangle \,=
$$

\be
\Big (\frac{\mu}{2 \pi} \Big )^2\,{\cal Z}_c^4 (\beta , \mu^\prime )\,
\lim_{\varepsilon^\pm ,\,\delta^\pm\,  \rightarrow 0}\,
\langle 0, \widetilde 0 \vert \,
\Big (\,W^\ast (x^\pm + \varepsilon^\pm; \beta )\,W (x^\pm; \beta )\,\Big )\,
\Big (\,W^\ast (y^\pm + \delta^\pm ; \beta )\,W ( y^\pm; \beta )\,\Big )\,\vert \widetilde 0, 0 \rangle\,.
\ee

\no Since the superselection
rule is identically satisfied, the dependence on $\mu$ and $\mu^\prime$ 
cancels and we get,

$$
\langle 0 (\beta ) \vert J^\pm (x)\,J^\pm (y ) \vert 0(\beta ) \rangle\,=
$$

$$
\Big (\frac{\mu}{2 \pi} \Big )^2\,\lim_{\varepsilon^\pm ,\, \delta^\pm\, \rightarrow 0}\,\Big (\,\frac{-\,1}{\mu^2\,\varepsilon\,\delta}\,\Big )\,
\langle 0, \widetilde 0 \vert \,\Big (\, 1 - 2\,i\,
\sqrt \pi\,\varepsilon^\pm\,\partial_{x^\pm}\, \phi (x^\pm ; \beta  )\,
\Big )\,\Big (\,1 - 2 \,i\,\sqrt \pi\,\delta^\pm\,
\partial_{y^\pm}\,\phi (y^\pm ; \beta )\,\Big )\,\vert \widetilde 0, 0  \rangle
$$

\be\label{sd}
=\,\frac{1}{\pi} \,\langle 0, \widetilde 0 \vert \,\partial^{x^\pm}\, 
\phi (x^\pm ; \beta )\,\partial_{y^\pm}\,\phi (y^\pm ; \beta )\,)
\,\vert \widetilde 0, 0 \rangle\,-\,\frac{1}{4\,\pi^2\,\varepsilon^\pm\,
\delta^\pm}\,.
\ee

\no The singular term appearing in (\ref{sd}) can be removed by redefining
the point-splitting by subtracting  the 
contribution coming from the vacuum 
expectation value of the individual current operators, i. e.,

$$
\langle 0 (\beta ) \vert J^\pm (x)\,J^\pm (y ) \vert 0 (\beta ) \rangle\,=
$$

$$
\lim_{\varepsilon ,\,\delta^\pm \,\rightarrow^\pm 0}\,
\Bigg \{\,\langle 0 (\beta ) \vert\,\Big (\,\psi^\dagger (x^\pm + \varepsilon^\pm)\,
\psi (x^\pm)\,\Big )\,\Big (\,\psi^\dagger (y^\pm + \delta^\pm )\,
\psi (y^\pm)\,\Big )\,\vert 0 (\beta ) \rangle \,-
$$

\be
-\,\langle 0 (\beta ) \vert\,\psi^\dagger (x^\pm + \varepsilon^\pm)\,
\psi (x^\pm)\,\vert 0 (\beta ) \rangle\,\langle 0 (\beta ) \vert\,
\psi^\dagger (y^\pm + \delta^\pm )\,
\psi (y^\pm)\,\vert 0 (\beta ) \rangle \,\Bigg \}\,.
\ee
 
\no The same for $\widetilde J^\pm (x^\pm ; \beta )$.

\subsection{Bosonized Lagrangian}

The total Lagrangian density of the free massless Dirac thermofield,

\be\label{tfld}
{\cal L}_T = i\, \overline{\Psi} (x ; \beta ) \gamma^\mu \partial_\mu 
\Psi (x ; \beta ) = {\cal L} - \widetilde{\cal L} =
i\, \bar \psi (x ; \beta ) \gamma^\mu \partial_\mu \psi (x ; \beta )\,
-\,\Big ( - 
i\, \overline{\widetilde \psi} (x ; \beta ) \gamma^\mu \partial_\mu 
\widetilde \psi (x ; \beta ) \Big )\,,
\ee
 
\no is equivalently given by,

\be\label{bl}
{\cal L}^{bos}_T = \frac{1}{2}\,:\partial_\mu \phi (x ; \beta )\partial^\mu \phi (x ; \beta ):\,
-\,
\frac{1}{2}\,:\partial_\mu \widetilde \phi (x ; \beta )\partial^\mu \widetilde \phi (x ; \beta ):\,.
\ee
 
\no The demonstration follows along the same lines as that given in Appendix E 
in \cite{AAR} for the zero temperature case. To this end, we must 
consider the Wilson short distance expansion for the total Hamiltonian
density

\be\label{th}
{\cal H}_T = \,i\,\overline\Psi\,\gamma^1\,\partial_1\,\Psi = 
{\cal H} - \widetilde{\cal H} = i \bar 
\psi \gamma^1 \partial_1 \psi - \Big ( - i \overline{\widetilde \psi} \gamma^1 \partial_1 
\widetilde \psi \Big ) + h. c.\,\,.
\ee

\no Since the short-distance 
behavior of the thermofield is the same as that
of zero temperature, we find

$$
{\cal H} (x; \beta ) = \frac{1}{2}\,\Big ( \bar \psi (x + \varepsilon ; \beta ) i \gamma^1 \partial_1
\psi (x ; \beta ) + h. c. \Big )\,-\,(V. E. V.)_\beta
$$

\be
=\,\frac{1}{2}\Big [\,:\Big (\partial_0 \phi (x ; \beta ) \Big )^2: 
+ :\Big (\partial_1 \phi (x ; \beta ) \Big )^2: \Big ]\,.
\ee 
 
\no For $\widetilde{\cal H}$ we find, 

$$
\widetilde{\cal H} (x ; \beta ) =\,-\,
\frac{1}{2}\,\Big ( \overline{\widetilde\psi} (x + \varepsilon ; \beta ) i \gamma^1 \partial_1
\widetilde\psi (x ; \beta ) + h. c. \Big )\,-\,(V. E. V.)_\beta
$$

\be\label{thamilt}
=\,-\,e^{\textstyle\,-\,i\,\pi}\,\frac{1}{2}\Big [\,:\Big ( \partial_0 \widetilde\phi (x ; \beta )
\Big )^2: 
+ :\Big (\partial_1 \widetilde\phi (x ; \beta ) \Big )^2: \Big ]\,.
\ee

\no The phase factor $e^{ - i \pi}$ in (\ref{thamilt}), that cancels the global
minus sign, comes from the phase (\ref{phase}) associated with the Wick 
exponential $\widetilde W (x ; \beta )$. The total Hamiltonian 
density ${\cal H}_T = {\cal H} - \widetilde{\cal H}$ 
corresponds to the total Lagrangian density (\ref{bl}) for 
the free massless scalar thermofield.



\section{Massless Thirring Model}

\setcounter{equation}{0}

In this section we illustrate the use of the thermofield bosonization
by solving the massless Thirring model, which is defined by
the Lagrangian density

\be
{\cal L}_T = i \bar \psi \gamma^\mu \partial_\mu \psi + \frac{1}{2} g\,(\bar \psi \gamma^\mu \psi ) ( \bar \psi \gamma_\mu \psi )
- \Bigg ( - i \widetilde{\bar\psi} \gamma^\mu \partial_\mu \widetilde\psi + \frac{1}{2} g\,(\widetilde{\bar\psi} \gamma^\mu \widetilde\psi ) 
( \widetilde{\bar\psi} \gamma_\mu \widetilde\psi ) \Bigg )\,.
\ee
 
To begin with, let us consider the operator solution of the 
(zero temperature) massless Thirring model,

\be\label{ost}
\psi (x) = \,\frac{1}{\sqrt \pi}\,\Big ( \mu \Big )^{\,\frac{1}{4 \pi} 
(\alpha^2 + \delta^2 )}\,
\,: e^{\,i\,\Big ( \alpha\,\gamma^5\,\varphi (x) + \delta\, \phi (x)\Big )}:\,,
\ee

\be
\widetilde\psi (x) = \,\frac{1}{\sqrt \pi}\,\Big ( \mu \Big )^{\,\frac{1}{4 \pi} 
(\alpha^2 + \delta^2 )}\,
\,: e^{\,-\,i\,\,\Big ( \alpha\,\gamma^5\,\widetilde\varphi (x) + \delta\, 
\widetilde\phi (x)\Big )}:\,,
\ee

\no  where $\varphi$ ($\widetilde\varphi$) is the dual 
of $\phi$ ($\widetilde\phi$) and $\mu$ is the infrared regulator
of the two-dimensional free massless scalar field. The $\mu$-dependent factor 
in front of (\ref{ost}) is needed in order to obtain in the IR limit $\mu \to 0$
well defined 
Wick exponentials living in a positive-metric  Hilbert space. The 
corresponding thermofield is given by,

\be
\psi (x ; \beta ) =  \frac{1}{\sqrt \pi}\,
\Big ( \mu \Big )^{\,\frac{1}{4 \pi} 
(\alpha^2 + \delta^2 )}\,
{\mathsf{Z}} (\beta , \mu^\prime )\,: e^{\,i\,\Big ( \alpha\,
\gamma^5\,\varphi (x ; \beta ) + \delta \phi (x ; \beta ) \Big )}:\,,
\ee

\no where

\be
{\mathsf{Z}} (\beta , \mu^\prime ) = e^{ \,-\,\frac{\alpha^2 + \delta^2}{\pi}\,\mathit{z} (\beta , \mu^\prime )\,-\,\alpha \delta \gamma^5\,
\Big ( [ \widetilde \varphi^{(+)}_s (x + \varepsilon ; \beta )\,,\,\widetilde \phi^{(-)}_s (x ; \beta ) ]\,+\,
 [ \widetilde \phi^{(+)}_s (x + \varepsilon ; \beta )\,,\,\widetilde \varphi^{(-)}_s (x ; \beta ) ] \Big )   }\,,
\ee
 
\no and similarly for $\widetilde\psi (x ; \beta )$. We shall consider in 
particular

\be
\delta = \frac{\pi}{\alpha}\,,
\ee

\no in order to obtain in the zero temperature limit a canonical 
fermion field with Lorentz spin $s = \frac{1}{2}$,

\be
s = \frac{\alpha\,\delta}{2 \pi} = \frac{1}{2}\,.
\ee

Taking into account that the short-distance behavior of the thermofield
two-point function is the same as that of zero temperature, we use the 
Mandelstam prescription \cite{Mandelstam, AAR} and define the current through the
point-splitting,

\be
{\cal J}^\mu (x ; \beta ) = 
\lim_{\varepsilon = (0 , \varepsilon^1)\atop{\varepsilon^1 \rightarrow 0}}\,
{\cal F} (\varepsilon )\,\Bigg ( \delta^0_\mu + \frac{\alpha^2}{\pi}
\delta^1_\mu\,\Bigg )\Big [ \,\bar \psi (x + \varepsilon ; \beta )\,
\gamma^\mu\,\psi (x ; \beta ) -
\langle 0 , \widetilde 0 \vert\,
\bar \psi (x + \varepsilon ; \beta )\,
\gamma^\mu\,\psi (x ; \beta )\vert \widetilde 0 , 0 \rangle \Big ] ,
\ee

\no where,

\be
{\cal F} (\varepsilon ) = 
\Big ( - \mu^2\,\varepsilon^2 \Big )^{\frac{\alpha^2 + \delta^2 - 2 \pi}{4 \pi}}\,.
\ee

\no One finds,

\be\label{tc}
{\cal J}^\mu (x ; \beta ) = -\,\frac{\alpha}{2\pi}\,
\epsilon^{\mu\,\nu}\,\partial_\nu\,\varphi (x ; \beta )\,,
\ee

\be
\widetilde{\cal J}^\mu (x ; \beta ) = -\,\frac{\alpha}{2\pi}\,
\epsilon^{\mu\,\nu}\,\partial_\nu\,\widetilde\varphi (x ; \beta )\,.
\ee

The quantum equations of motion are given in terms of the symmetrized form
of the interaction term of the Thirring current with the Fermi thermofields by,

\be
i \gamma^\mu \partial_\mu \psi (x ; \beta ) = 
- g \Big \{ {\cal J}_\mu (x ; \beta ) \gamma^\mu\,,\,\psi (x ; \beta ) \Big \}\,,
\ee

\be
i \gamma^\mu \partial_\mu \widetilde \psi (x ; \beta ) = 
  g \Big \{ \widetilde{\cal J}_\mu (x ; \beta ) \gamma^\mu\,,\,\widetilde\psi (x ; \beta ) \Big \}\,,
\ee

\no with the following identification 

\be
\alpha^2 = \frac{4 \pi}{1 - \frac{g}{\pi}}\,.
\ee

The unmixed correlation functions for the thermofield spinor 
components $\psi_\alpha (x ; \beta )$ are given by,

$$
\langle 0 , \widetilde 0 \vert \psi_\alpha (x_1 ; \beta ) \cdots 
\psi_\alpha (x_n ; \beta )\, \psi_\alpha^\dagger (y_1 ; \beta ) 
\cdots \psi_\alpha^\dagger (y_n ; \beta ) \vert \widetilde 0 , 0 \rangle 
=\,( 2 \pi )^{-n}\,
\Big ( i \Big )^{\,-\,
\frac{n}{2 \pi}\,\Big ( \delta^2 + \frac{\pi}{\delta^2} \Big )}
\times
$$

$$
\prod_{i < j}^n\,
\Bigg [ \frac{\beta}{\pi}\,\sinh\,\frac{\pi}{\beta}\,
\Big ( x^+_i - x^+_j - i \epsilon \Big )
\Bigg ]^{\frac{1}{4 \pi}\,
\Big ( \delta^2 + \frac{\pi^2}{\delta^2} + 2 \pi \,\gamma^5_{\alpha \alpha}
 \Big )}\,
\prod_{i , j}^n\,
\Bigg [ \frac{\beta}{\pi}\, \sinh\,\frac{\pi}{\beta}\,
\Big ( x^-_i - x^-_j - i \epsilon \Big )\Bigg ]^{\frac{1}{4 \pi}\,
\Big ( \delta^2 + \frac{\pi^2}{\delta^2} - 2 \pi\, \gamma^5_{\alpha \alpha}
 \Big )}\,\times 
$$

$$
\prod_{i < j}^n\,
\Bigg [ \frac{\beta}{\pi}\,\sinh\,\frac{\pi}{\beta}\,
\Big ( y^+_i - y^+_j - i \epsilon \Big )\Bigg ]^{\frac{1}{4 \pi}\,
\Big ( \delta^2 + \frac{\pi^2}{\delta^2} + 2 \pi \,
\gamma^5_{\alpha \alpha} \Big )}\,
\prod_{i , j}^n\,
\Bigg [\frac{\beta}{\pi} \sinh\,\frac{\pi}{\beta}\,
\Big ( y^-_i - y^-_j - i \epsilon \Big )\Bigg ]^{\frac{1}{4 \pi}\,
\Big ( \delta^2 + \frac{\pi^2}{\delta^2} - 2 \pi\,\gamma^5_{\alpha \alpha}
 \Big )}\,\times 
$$

$$
\prod_{i , j}^n\,
\Bigg [ \frac{\beta}{\pi}\,\sinh\,\frac{\pi}{\beta}\,
\Big ( x^+_i - y^+_j - i \epsilon \Big )\Bigg ]^{\,-\,\frac{1}{4 \pi}\,
\Big ( \delta^2 + \frac{\pi^2}{\delta^2} + 2 \pi\,\gamma^5_{\alpha \alpha}
 \Big )}\,
\prod_{i , j}^n\,
\Bigg [ \frac{\beta}{\pi}\, \sinh\,\frac{\pi}{\beta}\,
\Big ( x^-_i - y^-_j - i \epsilon \Big )\Bigg ]^{\,-\,\frac{1}{4 \pi}\,
\Big ( \delta^2 + \frac{\pi^2}{\delta^2} - 2 \pi\,
\gamma^5_{\alpha \alpha} \Big )}\,.
$$

\no which represents the generalization of the $T = 0$ result to
$T \neq 0$.

\section{Conclusion}

Using the framework of Thermofield Dynamics \cite{Das} we have 
shown how the familiar $T=0$ bosonization formulae for fermions 
in 1+1 dimensions generalize in a natural way to the case of 
non-zero temperature. Despite the appearance of now two 
infrared regulating parameters it was shown that the superselection 
rule already familiar from the $T=0$ case guaranteed the 
independence of physical quantities of both these parameters. The 
``Mandelstam representation" for thermal fermions now allows us to
obtain in a compact way the exact and complete solution of a 
number of integrable models involving massless fermions in 1+1 dimensions.

The next step would now address the problem of massive fermions in
1+1 dimensions. Do we again have an equivalence between the 
Massive Thirring model and the sine-Gordon theory ? We expect that this is
indeed the case, since such an equivalence has been demonstrated for 
correlators of fermion and sine-Gordon thermal fields in the
imaginary time formalism \cite{Delepine}. Taking a Mandelstam representation
in terms of a thermal sine-Gordon field as a starting point, can we 
prove, in a way analogous to what was done in the
$T=0$ massive Schwinger model \cite{RS}, that the 
bosonized thermal Dirac field satisfies the free massive Dirac equation ? These
and other questions will be subject of future investigations. 

\centerline{Acknowledgments}

We are grateful to the CNPq-DAAD scientific exchange program which makes this
collaboration possible. 

\appendix{{\centerline{\bf{Appendix}}}}
\renewcommand{\theequation}{{A}.\arabic{equation}}
\setcounter{equation}{0}

The purpose of this appendix is to show that the theory of the
free massless scalar thermofield can be constructed as the zero mass limit
of the free massive scalar thermofield theory. To begin with, let 
us consider the free massive scalar thermofield,

$$
\Sigma (x ; \beta ) = U [ \theta_B (\beta ) ]\,\Sigma (x)\,U [ \theta_B (\beta ) ]
=
$$

$$
\frac{1}{2 \sqrt \pi}\,\int_{- \infty }^{+ \infty}\,\frac{d p}{\sqrt{p^2 + m^2}}\,
\Big \{f_p (x) \Big ( a (p) \cosh \theta_B (p ; \beta) - \widetilde a^\dagger (p) \sinh 
\theta_B (p ; \beta) \Big ) 
$$

\be
+
f^\ast_p (x) \Big ( a^\dagger (p) \cosh \theta_B (p ; \beta) - \widetilde a(p) \sinh 
\theta_B (p ; \beta) \Big )\Big \}\,.
\ee

\no The thermal two-point function is given by,

\be\label{m2pf1}
\langle 0,\widetilde 0 \vert \Sigma (x ; \beta ) \Sigma (y ; \beta ) 
\vert \widetilde 0 , 0 \rangle = D^{(+)}_o (x - y ; m ) +
{\cal I} (x - y ; \beta , m)\,,
\ee

\no where $D^{(+)}_o (x - y ; m)$ is the corresponding zero temperature two-point 
function,

\be\label{i1}
D^{(+)}_o (x - y ; m) =
\frac{1}{4\pi}\,
\int_{0}^{\infty}\,\frac{d p}{\sqrt{p^2 + m^2}}\,\Big \{
e^{ - i \Big [ (x^0 - y^0) \sqrt{p^2 + m^2} +  (x^1 - y^1) p  \Big ]} +
e^{ - i \Big [ (x^0 - y^0) \sqrt{p^2 + m^2} -  (x^1 - y^1)p \Big ]}\Big \}\,,
\ee

\no and the last integral in (\ref{m2pf1}) can
be written as,

$$
{\cal I} (x - y ; \beta , m) = 
\frac{1}{2\pi}\,
\int_{0}^{\infty}\,\frac{d p}{\sqrt{p^2 + m^2}}\,N_B (w , \beta )\,
\Big \{\cos \Big [ (x^0 - y^0) \sqrt{p^2 + m^2} + (x^1 - y^1) p \Big ]\,
$$

\be\label{i2}
+ \cos \Big [ (x^0 - y^0) \sqrt{p^2 + m^2} -  (x^1 - y^1) p \Big ] \Big \}\,,
\ee

\no where

\be
N_B (w ,\beta ) = \frac{1}{e^{\beta w} - 1}\,=\,\frac{1}{e^{\beta \sqrt{p^2 + m^2}} - 1}\,.
\ee

\no As long as $m \neq 0$ the integrals (\ref{i1}) and (\ref{i2}) 
are not infrared divergent. Thus, before considering the zero mass limit
an infrared cut-off $\mu > 0$ must be introduced. In the 
limit $m \rightarrow 0$, $N_B (w ,\beta ) \rightarrow N_B (p ; \beta )$ and
we obtain,

\be
D^{(+)} ( x - y ; m , \beta )_{_{m \rightarrow 0}} =
D^{(+)}_o (x - y ) +
\frac{1}{2\pi}\,
\int_{\mu}^{\infty}\,\frac{d p}{p}\,N_B (p ;\beta )\,
\Big [ \cos p (x^+ - y^+) + \cos p (x^- - y^- ) \Big ] \,,
\ee

\no which is the two-point function of the free massless
thermofield given by Eq. (\ref{2pf}). By considering the free massless
scalar thermofield theory as the zero mass limit of the free massive
scalar thermofield theory, the infrared regulator $\mu$ of 
the zero temperature two-point function should be identified with the
infrared cut-off $\mu^\prime$ of the temperature dependent contribution
in (\ref{2pf}).

\end{document}